\PassOptionsToPackage{table}{xcolor}  % o las opciones que necesites
\documentclass{article}

\usepackage{arxiv}
\usepackage{setspace}
\usepackage[utf8]{inputenc} % allow utf-8 input
\usepackage[T1]{fontenc}    % use 8-bit T1 fonts
\usepackage{url}            % simple URL typesetting
\usepackage{booktabs}       % professional-quality tables
\usepackage{amsfonts}       % blackboard math symbols
\usepackage{nicefrac}       % compact symbols for 1/2, etc.
\usepackage{microtype}      % microtypography
\usepackage{lipsum}
\usepackage{graphicx}
\graphicspath{ {./images/} }

\usepackage{amsmath}
\DeclareMathOperator{\logit}{logit}
\usepackage{multirow}
\usepackage{makecell}
\usepackage{xcolor}

\usepackage[super,numbers,sort&compress]{natbib}
\title{A multivariate spatial model for ordinal survey-based data}

\author{
 Miguel Ángel Beltrán-Sánchez \orcidlink{0000-0001-9450-2973} \\
  Department of Statistics and Operations Research \\
  University of Valencia, Burjassot (Valencia), Spain \\
  \texttt{angel.beltran@uv.es} \\
   \And
 Miguel Ángel Martínez-Beneito \orcidlink{0000-0001-8406-8050} \\
  Department of Statistics and Operations Research \\
  University of Valencia, Burjassot (Valencia), Spain \\
  \texttt{miguel.a.martinez@uv.es} \\
  \And
 Ana Corberán-Vallet \orcidlink{0000-0002-1091-9534} \\
  Department of Statistics and Operations Research \\
  University of Valencia, Burjassot (Valencia), Spain \\
  \texttt{ana.corberan@uv.es} \\
}

\begin{document}
\maketitle
\begin{abstract}
Health surveys provide valuable information for monitoring population health, identifying risk factors and informing public health policies. Most of the questions included are coded as ordinal variables and organized into thematic blocks. Accordingly, multivariate modeling provides a natural framework for considering these variables as true groups, thereby accounting for potential dependencies among the responses within each block. In this paper, we propose a multivariate spatial analysis of ordinal survey-based data. This multivariate approach enables the joint analysis of sets of ordinal responses that are likely to be correlated, accounting for individual-level effects, while simultaneously improving the estimation of the geographical patterns for each variable and capturing their interdependencies. We apply this methodology to describe the spatial distribution of several mental health indicators from the Health Survey of the Region of Valencia (Spain) for the year 2022. Specifically, we analyze the block of questions from the 12-item General Health Questionnaire included in the survey.
\end{abstract}

% keywords can be removed

\section{Introduction} \label{sec:introduction}

Small-area spatial analyses are a valuable epidemiological tool to provide accurate estimates for regional indicators and understand spatial patterns. Multiple measurements for diseases, risk factors or health indicators are often recorded at each areal unit. On those occasions, multivariate spatial models are able to account for spatial dependence between locations as well as dependence among the different response variables.

In the context of multivariate disease mapping, Martínez-Beneito\cite{MartinezBeneito2013} proposed a unifying modeling framework that subsumes most of the statistical models in the literature addressing this problem, specifically those based on multivariate conditional autoregressive (MCAR) distributions. This proposal is able to reproduce both separable and non-separable covariance structures and can accommodate different spatial correlation structures within diseases. This approach was later on reformulated in order to develop a computationally more efficient form that can handle a larger number of diseases. \cite{BotellaRocamora2015} By using information from correlated diseases, the multivariate model allows for borrowing strength across outcomes, thereby providing more reliable estimates of the spatial risk surfaces as compared to traditional univariate models, especially for low-mortality diseases and for those exhibiting similar geographical patterns. It can also help us find common risk factors between diseases. This modeling approach has been applied, for instance, to study the spatial distribution of mortality in urban areas through a comprehensive analysis of 16 causes of death across several Spanish cities\cite{MartinezBeneito2021}, or to cancer mortality data at the municipality level in Spain, through an extension that partitions the spatial domain into smaller subregions. \cite{Vicente2023}

Unfortunately, disease registries and administrative health databases capture only a small fraction of the information relevant to public health research. Many relevant risk factors and health indicators, such as health behaviors or mental health conditions, are often missing from these sources, making periodic health surveys a crucial complement to health surveillance. However, survey-based analyses are often limited to basic descriptive summaries, including frequency tables across sex, age, education or administrative regions. If we were interested in assessing statistically significant differences between demographic groups, identifying relevant associations among health indicators within specific thematic blocks, or estimating the geographical pattern of each response variable, it would be convenient to consider multivariate spatial models for analyzing survey-based data. Despite its relevance, the existing literature on multivariate spatial modeling with survey data is relatively limited. Therefore, it is a rich field of study.

Most existing studies on survey-based data focus on binary responses that are modeled using either Bernoulli or Binomial (for data aggregated in small areas) distributions. In this context, for the multivariate analysis of the National Health Survey of Chile, Lawson et al.\cite{Lawson2020} proposed a multi-scale joint model. This model allows for correlation between areas and captures the links between individual outcomes. Spatial correlation was modeled using a Markov Random Field (MRF). Yet, different spatial patterns were assumed for each response variable. In parallel, a hierarchical Bayesian spatial shared component model was proposed to assess the co-occurrence of acute respiratory infection, diarrhea and stunting in children aged below 5 years in Nigeria. \cite{Orunmoluyi2022} The model allows for the estimation of both an underlying disease-overall spatial risk pattern and disease-specific spatial components. Bayesian spatial multivariate shared component modeling was also used to assess geographical and socio-demographic factors influencing risky sexual behaviors among young adults in Nigeria. \cite{Seidu2024} More recently, Cassy and Manda\cite{Cassy2024} proposed a pseudo-likelihood approach that combines a weighted likelihood, which integrates the normalized design weights, with the shared spatial component model to estimate the geographic distribution of childhood fever and diarrhea prevalence in Mozambique. Finally, Yu et al.\cite{HanjunYu2025} proposed a Bayesian hierarchical bivariate spatial model for small-area estimation of proportions using MCAR distributions applied to health survey data from China.

Despite the assumption of Bernoulli/Binomial distributions in the mentioned works, it is important to emphasize that health surveys typically include questions with multiple response options, allowing respondents to select the one that best reflects their situation. These options are usually ordered or ranked, and so health indicators are often coded as ordinal variables. Recently, a Bayesian hierarchical individual-level model has been proposed for the spatial analysis of ordinal data from health surveys. \cite{BeltranSanchez2024} This work integrates several statistical research areas, namely spatial statistics, the analysis of ordinal data and the analysis of survey data collected under complex sampling designs, in which all design variables are assumed to be known. Yet, as a univariate framework, it fails to capture the potential dependence between the different outcomes included in a particular thematic block, such as mental health conditions, physical limitations, social support needs or health behaviors. To our knowledge, there are currently no multivariate modeling approaches that jointly handle ordinal survey-based data while also accounting for small-area spatial dependence.

The main objective of this work is to extend the previously proposed model for survey-based ordinal data to a multivariate setting. This multivariate extension enables the joint analysis of sets of ordinal health indicators that are likely to be correlated. In fact, the multivariate model allows us to estimate, and take advantage of, the correlation between the different geographical patterns. Specifically, by borrowing strength from related health data, the estimation of the question-specific spatial patterns is improved. In addition, ordinal survey-based data can be exploited more effectively by constructing synthetic measures, such as maps of the spatial patterns associated with each survey question and correlation matrices between health indicators at both respondent and area levels, providing richer insights and more concise summaries of the results. Ultimately, this methodology allows for the identification of relationships among health indicators within the thematic block and the extraction of valuable information that would otherwise remain hidden. As a result, relevant authorities can make more informed decisions or implement targeted interventions in areas exhibiting higher levels of risk, thereby maximizing the potential of health survey data.

This paper is organized as follows. Section~\ref{sec:methodology} reviews the individual-level model for small-area estimation of ordinal survey-based health indicators. Three multivariate modeling extensions are subsequently introduced. Section~\ref{sec:case-study} applies the methodology to describe the spatial distribution of several mental health indicators from the Health Survey of the Region of Valencia (Spain) for the year 2022 (HSRV2022). Finally, Section~\ref{sec:discussion} concludes with some final remarks and future work.

\section{Methodology} \label{sec:methodology}

In this section, we first provide an overview of the previously mentioned univariate individual-level model for small-area inference with ordinal data from health surveys. Subsequently, we introduce several multivariate extensions within this framework.

\subsection{Background} \label{sec:background}

A Bayesian hierarchical individual-level model for the analysis of spatial ordinal survey-based data has recently been proposed. \cite{BeltranSanchez2024} This novel approach jointly incorporates spatial dependence and sampling design information in a unified modeling stage. By conditioning the individual-level ordinal response on all the design variables, it allows for the estimation of finite population quantities, such as the proportion of individuals in each health status category at any desired hierarchical level (e.g., small-area level), through the post-stratification of the model results.

Let $Y_i \in \{1, \dots, J\}$ be an ordinal response variable for the $i$-th individual of the population of interest ($i = 1, \dots, N$). Suppose the study region is divided into $M$ small areas, being $\{N_m\}_{m = 1}^M$ their known population sizes, so that the total population size is $N = \sum_{m=1}^M N_m$. A sample of size $n_m$ is drawn from area $m$ according to a known sampling scheme (e.g., simple random sampling, stratified or cluster sampling). Consequently, the total sample size is $n = \sum_{m = 1}^M n_m < N$, meaning that only a subset of the values for $Y_i$ is observed; that is, those corresponding to the $n$ survey respondents. Without loss of generality, we assume that the first $n$ individuals in the population correspond to the survey respondents, and the remaining $N - n$ are all unobserved. Let $\boldsymbol{y} = (y_1, \dots, y_n)^{T}$ denote the vector containing the responses and $\boldsymbol{x}_i = (z_i, m_i)^{T}$, $i = 1, \dots, n$, represent the covariate vector for the $i$-th respondent, where $z_i \in \{1, \dots, Z\}$ is the value of a categorical covariate (e.g., stratum in a stratified sampling design) and $m_i \in \{1, \dots, M\}$ indicates the small area to which the respondent belongs. It is important to ensure that the covariate vector $\boldsymbol{x}_i$ includes all the design variables, what makes the design ignorable given those variables; otherwise, the resulting inferences may not be valid for the target population. \cite{Little2003, Gelman2013}

The likelihood for the $i$-th respondent is specified as follows: 
\begin{equation*}
y_i|\boldsymbol{\pi}(\boldsymbol{x}_i) \sim Categorical(\boldsymbol{\pi}(\boldsymbol{x}_i)),
\end{equation*} 
where $\boldsymbol{\pi}(\boldsymbol{x}_i) = (\pi_1(\boldsymbol{x}_i), \pi_2(\boldsymbol{x}_i), \dots, \pi_J(\boldsymbol{x}_i))^{T}$ is the vector of the probabilities for the different categories given $\boldsymbol{x}_i$; that is, $\pi_j(\boldsymbol{x}_i) = P(Y_i = j | \boldsymbol{x}_i)$ for $j = 1, \dots, J$. Since $Y_i$ is ordinal, modeling cumulative probabilities instead of category-specific ones is often more appropriate. \cite{McCull1980, Congdon2005} Let $\gamma_j(\boldsymbol{x}_i) = \sum_{r = 1}^j \pi_r(\boldsymbol{x}_i)$ denote the cumulative probability of categories $1$ to $j$ given $\boldsymbol{x}_i$ for $j = 1, \dots, J$. The first $J - 1$ cumulative probabilities are modeled as follows: 
\begin{equation*}
\logit (\gamma_j(\boldsymbol{x}_i)) = \kappa_j + \alpha_{z_i} + \theta_{m_i},
\end{equation*} 
where $\kappa_1 < \kappa_2 < \dots < \kappa_{J-1}$, and $\{\alpha_z\}_{z = 1}^Z$ and $\{\theta_m\}_{m = 1}^M$ are intended to model the effects of the covariates and locations, respectively. In this proposal, non-informative priors are specified for both the cut points (intercepts $\{\kappa_j\}_{j = 1}^{J - 1}$) and the fixed effects associated with the categorical covariate $\{\alpha_z\}_{z = 1}^Z$; while spatial dependence is introduced for the geographical components $\{\theta_m\}_{m = 1}^M$ through a conditional autoregressive (CAR) prior. \cite{Leroux2000} Moreover, it is possible to allow the categorical covariate of interest to have different effects for each level of the cumulative probabilities. This can be achieved by letting the intercepts to vary for each category $j$ and each value of the categorical covariate: 
\begin{equation}
\logit (\gamma_j(\boldsymbol{x}_i)) = \kappa_{z_i, j} + \theta_{m_i}.
\label{eq:extension}
\end{equation}
In summary, the proposed methodology allows to analyze ordinal survey-based data at the individual level, estimating the effect of covariates associated with the outcome, while simultaneously accounting for spatial dependence and the data collection scheme at a single modeling stage. In addition, estimates can be made at either the individual level or aggregated up to any desired level of the model hierarchy via the multilevel regression with post-stratification (MRP) approach. \cite{Gelman1997, Park2004, Park2006} The MRP process allows for transforming the model estimates, which correspond to superpopulation quantities, to the population level, thereby obtaining estimates of finite population quantities.

\subsection{Spatially structured multivariate modeling} \label{sec:methods}

From now on, we detail three multivariate modeling extensions of the previous model, ordered from the simplest to the most complex one.

Let us assume that we were interested in a block of questions in a survey composed of $K$ ordinal variables. Without loss of generality, we assume they all have the same number $J$ of categories, ordered from better to worse health status. Let $Y_{ik} \in \{1,\dots,J\}$ be the $k$-th ordinal response variable that quantifies a health indicator for the $i$-th individual of the population of interest, $i = 1, \dots, N$. We assume that the first $n$ individuals in the population correspond to the survey respondents, and the remaining $N - n$ are all unobserved. Let $\boldsymbol{y}_i = (y_{i1}, \dots, y_{iK})^{T}$ denote the vector containing the observed values for the $i$-th respondent. Let us now assume that $N_{\cdot}$ and $n_{\cdot}$ represent, respectively, the population and sample size corresponding to their subscript. We will focus on the case of a single categorical covariate with $Z$ levels, as for example the stratum in a stratified sampling design or an auxiliary variable unrelated to the sampling design that could be used to improve the proportion estimates. In that case, $\boldsymbol{x}_i = (z_i, m_i)^{T}$ is the covariate vector that indicates the level of the categorical covariate $z_i \in \{1, \dots, Z\}$ and the small area $m_i \in \{1, \dots, M\}$ to which the $i$-th respondent belongs.

The following likelihood is assumed for the ordinal variables: 
\begin{equation}
y_{ik}|\boldsymbol{\pi}_k(\boldsymbol{x}_i) \sim Categorical(\boldsymbol{\pi}_k(\boldsymbol{x}_i)),
\label{eq:likelihood}
\end{equation} 
where $\boldsymbol{\pi}_k(\boldsymbol{x}_i) = (\pi_{1k}(\boldsymbol{x}_i), \dots, \pi_{Jk}(\boldsymbol{x}_i))^{T}$ denotes the vector of the probabilities for the different categories of the $k$-th ordinal variable conditional on $\boldsymbol{x}_i$; that is, $\pi_{jk}(\boldsymbol{x}_i) = P(Y_{ik} = j | \boldsymbol{x}_i), \ j = 1, \dots, J$. Let $\gamma_{jk}(\boldsymbol{x}_i)$ denote the cumulative probability of categories $1$ to $j$ for the $k$-th response given $\boldsymbol{x}_i$; that is: 
\begin{align}
\gamma_{jk}(\boldsymbol{x}_i) &= P(Y_{ik} \leq j | \boldsymbol{x}_i) \label{eq:cumulative-probs} \\  &= \pi_{1k}(\boldsymbol{x}_i) + \pi_{2k}(\boldsymbol{x}_i) + \dots + \pi_{jk}(\boldsymbol{x}_i), \notag
\end{align} 
where $0 < \gamma_{1k}(\boldsymbol{x}_i) < \gamma_{2k}(\boldsymbol{x}_i) < \dots < \gamma_{J-1, k}(\boldsymbol{x}_i) < 1$. The modeling of the previous cumulative probabilities with different effects in the linear predictor will define distinct proposals for the spatial multivariate analysis of ordinal variables.

\subsubsection{Multivariate modeling with independent responses} \label{sec:indep}

In this first proposal, we assume independence among the $K$ ordinal responses within the corresponding block of survey questions. This ``naive'' proposal is introduced for comparison with the rest of models. Assuming the likelihood in Expression~\ref{eq:likelihood}, the first $J-1$ cumulative probabilities defined in Expression~\ref{eq:cumulative-probs} are modeled as follows: 
\begin{equation}
\logit (\gamma_{jk}(\boldsymbol{x}_i)) = \kappa_{jk} + \alpha_{z_i k} + \theta_{m_i k},
\label{eq:model-indep}
\end{equation} 
where $\kappa_{1k} < \kappa_{2k} < \dots < \kappa_{J-1, k}$ for $k = 1, \dots, K$. The collection of intercepts $\{\kappa_{j k}\}_{j = 1}^{J-1}$ corresponds to cut points that break the standard logistic probability density function for the $k$-th response into $J$ pieces, representing the average value of the associated cumulative probabilities on the logit scale. \cite{Congdon2005, Faraway2016, Dobson2018} Parameters $\{\alpha_{z k}\}_{z = 1}^{Z}$ are fixed effects associated with the categorical covariate for the $k$-th ordinal variable. Finally, the set $\{\theta_{m k}\}_{m = 1}^{M}$ represents a collection of spatial random effects for the $k$-th response; that is, each random vector $\boldsymbol{\theta}_{k} = (\theta_{1k}, \dots, \theta_{Mk})^{T}$ accounts for the spatial dependence corresponding exclusively to the $k$-th ordinal outcome. Recall that this model (as well as the following ones) can also be reformulated to allow the cut points to vary not only by $j$ and each variable $k$, but also with $z_i$, by considering $\kappa_{z_i j k}$ in the linear predictor (see Expression~\ref{eq:extension}).

For each response variable, the prior distributions for all the model parameters are specified as in the univariate approach. \cite{BeltranSanchez2024} In particular, we assume improper uniform priors for the fixed effects associated with the categorical covariate; that is, $\alpha_{1k}$ is fixed to $0$ in order to avoid confounding with the intercepts, and $\{p(\alpha_{z k})\}_{z=2}^Z \propto 1$ for $k = 1, \dots, K$. Instead of assuming corner constraints for $\{\alpha_{z k}\}_{z = 1}^{Z}$, as we have done, we could alternatively assume zero-sum constraints ($\alpha_{1k} = -\sum_{z = 2}^Z \alpha_{z k}$), depending on the convenience of each parameterization for each particular model.

Regarding the spatial random effects, we assume $\boldsymbol{\theta}_{k}$ to follow a Leroux CAR prior distribution\cite{Leroux2000}; that is, $\boldsymbol{\theta}_{k}|\sigma_k^2, \rho_k \sim LCAR(\sigma_k^2, \rho_k)$. We assign non-informative uniform priors to the associated hyperparameters: $\sigma_k \sim U(0, \infty)$ and $\rho_k \sim U(0, 1)$. To ensure identifiability, a zero-sum constraint is imposed on each of the $K$ spatial random vectors. Specifically, we require the sum of the spatial random effects across all sampled individuals to be zero; that is, $\sum_{i = 1}^n \theta_{m_i k} = 0$ for $k = 1, \dots, K$. In fact, those constraints are equivalent to assuming orthogonality restrictions between the $K$ spatial random vectors $\boldsymbol{\theta}_{k}$ and the $n$-dimensional vector of ones $\boldsymbol{1}_n$ corresponding to the intercepts. Note that these restrictions can alternatively be interpreted as weighted constraints, where the weights correspond to the number of respondents in each small area; that is, $\sum_{m = 1}^M n_{m}\theta_{m k} = 0$ for $k = 1, \dots, K$.

Finally, the prior distribution for each vector of cut points $\boldsymbol{\kappa}_{k} = (\kappa_{1k},\dots,\kappa_{J-1, k})^{T}$ is specified to ensure that all response categories have equal prior probabilities while respecting the order constraint $\kappa_{1k} < \kappa_{2k} < \dots < \kappa_{J-1, k}$. To achieve this, we introduce auxiliary vectors $\boldsymbol{\delta}_{k} = (\delta_{1k}, \dots, \delta_{Jk})^{T}$, such that $0 < \delta_{j k} < 1$, $j = 1, \dots, J$, and $\sum_{j = 1}^J \delta_{j k} = 1$, so that $\boldsymbol{\delta}_{k} \sim Dirichlet(\boldsymbol{1}_J)$, which results in a uniform distribution over the simplex. The vector $\boldsymbol{\kappa}_{k}$ is then defined as a monotonic transformation of the cumulative sums of the components of $\boldsymbol{\delta}_{k}$, for instance $\kappa_{j k} = \logit (\sum_{r = 1}^{j} \delta_{r k}), \ j = 1, \dots, J-1$. If direct modeling of the auxiliary vectors using a Dirichlet prior is not feasible, we could alternatively code that distribution resorting to a stick-breaking prior process. \cite{Sethuraman1994} Finally, if we model the cumulative probabilities using $\kappa_{z_i j k}$, then the prior distribution defined above is used for each value of the categorical covariate.

It is important to emphasize that this multivariate approch becomes equivalent to fitting $K$ separate univariate models, one for each of the ordinal variables, following the recent proposal by Beltrán-Sánchez et al.\cite{BeltranSanchez2024}.

\subsubsection{Multivariate modeling with correlated responses} \label{sec:corr}

In this second proposal, multivariate spatial dependence is introduced among the $K$ ordinal responses within the block of survey questions. Given the likelihood in Expression~\ref{eq:likelihood}, we model the cumulative probabilities defined in Expression~\ref{eq:cumulative-probs} as in Model~\ref{eq:model-indep}, but now $\theta_{m k}$'s are a collection of random effects whose joint distribution specifies the spatial dependency within-variables and the dependency between-variables at the small-area level; that is, the spatial random vectors $\boldsymbol{\theta}_{k}$ account for geographic dependencies across small areas for each health indicator, as well as cross-correlations among them.

In particular, spatial random effects are modeled according to the \textbf{M}--model framework introduced by Botella-Rocamora et al.\cite{BotellaRocamora2015}. Let $\boldsymbol{\Theta}$ be the $M \times K$ matrix that collects all $\theta_{m k}$'s. We assume: 
\begin{equation*}
\boldsymbol{\Theta} = \boldsymbol{\Phi} \boldsymbol{M},
\end{equation*} 
where $\boldsymbol{\Phi}$ is an $M \times K$ matrix of zero-mean latent Gaussian random effects composed of independent columns $\boldsymbol{\phi}_k$, each following a CAR prior distribution, and $\boldsymbol{M}$ is a nonsingular $K \times K$ matrix. Therefore, the columns of matrix $\boldsymbol{\Theta}$; that is, the vectors $\boldsymbol{\theta}_{k}$, are linear combinations of those underlying spatial random effects, with coefficients given by the corresponding columns of $\boldsymbol{M}$: 
\begin{equation*}
\boldsymbol{\theta}_{k} = \boldsymbol{\phi}_{1}M_{1k} + \boldsymbol{\phi}_{2}M_{2k} + \dots + \boldsymbol{\phi}_{K}M_{Kk},
\end{equation*} 
where $M_{ij}$ is the ($i$,$j$)-th entry in $\boldsymbol{M}$. We assume here each of the underlying spatial random effects to follow a Leroux CAR prior distribution with fixed variance equal to one, so that their variance depends on the values of $\boldsymbol{M}$; that is, $\boldsymbol{\phi}_{ k}|\rho_k \sim LCAR(\sigma_{k}^2 = 1, \rho_k)$. As for the entries in $\boldsymbol{M}$, we assume they are independent and identically distributed Gaussian random effects; that is, $M_{ij}|\sigma_{\boldsymbol{M}}^2 \sim N(0, \sigma_{\boldsymbol{M}}^2)$, and $\sigma_{\boldsymbol{M}} \sim U(0, \infty)$. Note that the assumption of a common scale parameter for the cells of $\boldsymbol{M}$ can be relaxed by using row or column variance-adaptive random effect \textbf{M}--models, yielding heteroscedasticity for the response variables. \cite{CorpasBurgos2019} To ensure identifiability, we impose the same $K$ zero-sum constraints described in Section~\ref{sec:indep} on the latent spatial random vectors $\boldsymbol{\phi}_{k}$; that is, $\sum_{m = 1}^M n_{m}\phi_{m k} = 0$ for $k = 1, \dots, K$. The matrix product $\boldsymbol{\Sigma}_{b} = \boldsymbol{M}^T \boldsymbol{M}$ represents the variance--covariance matrix of the spatial patterns across the different ordinal responses, thereby inducing cross-dependence between health indicators at the small-area level.

\subsubsection{Multivariate modeling with correlated responses and individual random effects} \label{sec:corrandire}

Finally, we also account for multivariate dependence among the $K$ ordinal variables provided by the same individual. When dealing with multivariate survey-based data, it is common to encounter correlations among responses that reflect unobserved individual-level characteristics. Ignoring such multivariate individual dependence may lead to misleading inferences, particularly at the small-area level. To properly capture this structure, we extend our modeling framework to incorporate respondent-specific effects. The inclusion of individual-level random effects is convenient for ensuring reliable correlations among geographical patterns and for filtering out potential individual-level correlations. Assuming the likelihood in Expression~\ref{eq:likelihood}, we model now the cumulative probabilities defined in Expression~\ref{eq:cumulative-probs} as:
\begin{equation}
\logit (\gamma_{jk}(\boldsymbol{x}_i)) = \kappa_{jk} + \alpha_{z_i k} + \theta_{m_i k} + \psi_{ik},
\label{eq:model-corr-ire}
\end{equation} 
where $\kappa_{1k} < \kappa_{2k} < \dots < \kappa_{J-1, k}$. The collection of cut points $\{\kappa_{j k}\}_{j = 1}^{J-1}$, fixed effects $\{\alpha_{z k}\}_{z = 1}^{Z}$ and spatial random effects $\{\theta_{m k}\}_{m = 1}^{M}$ for $k = 1, \dots, K$ have exactly the same meaning as those in the previously described model in Section~\ref{sec:corr}. The difference is the incorporation of a new collection of individual random effects (IREs) $\psi_{i k}$'s that captures both within-variables variability and between-variables dependency at the individual level; that is, the individual random vectors $\boldsymbol{\psi}_{k}$ account for respondent variability for each health indicator, as well as correlations among them.

With that goal, we model the IREs in a similar manner to the (multivariate) spatial random effects. Let $\boldsymbol{\Psi}$ be the $n \times K$ matrix that collects all $\psi_{i k}$'s. Following the \textbf{M}--model framework:
\begin{equation*}
\boldsymbol{\Psi} = \boldsymbol{\tilde{\Phi}} \boldsymbol{\tilde{M}},
\end{equation*} 
where $\boldsymbol{\tilde{\Phi}}$ is an $n \times K$ matrix of zero-mean latent Gaussian random effects and $\boldsymbol{\tilde{M}}$ a $K \times K$ matrix. Under this formulation, the matrix product $\boldsymbol{\tilde{\Sigma}}_{b} = \boldsymbol{\tilde{M}}^T \boldsymbol{\tilde{M}}$ defines the individual variance--covariance matrix across the ordinal variables, inducing multivariate dependence among them. As prior distributions, we assume that $\boldsymbol{\tilde{\phi}}_{ k} \sim N_n(\boldsymbol{0}_n, \boldsymbol{I}_n)$ for $k = 1, \dots, K$ where $\boldsymbol{I}_n$ is the identity matrix of size $n$, so that we assume independence between respondents, in contrast to the spatial term. Regarding the entries of the square matrix $\boldsymbol{\tilde{M}}$, they are assumed again to be independent and identically distributed Gaussian random effects; that is, $\tilde{M}_{ij}|\sigma_{\boldsymbol{\tilde{M}}}^2 \sim N(0, \sigma_{\boldsymbol{\tilde{M}}}^2)$. As for the spatial term, $\sigma_{\boldsymbol{\tilde{M}}}$ could also vary for each row or column of $\boldsymbol{\tilde{M}}$, thereby producing heteroscedastic processes. \cite{CorpasBurgos2019} 

\section{A case study} \label{sec:case-study}

In this section, we apply the methodology developed in Section~\ref{sec:methods} to analyze the block of questions from the 12-item General Health Questionnaire (GHQ--12)\cite{Goldberg1988} included in the HSRV2022. This block is designed to assess the mental health status of the population in the Region of Valencia (RV), one of the 17 Spanish regions, particularly focusing on emotional problems experienced during the past four weeks. All twelve items ($K = 12$) have four response levels ($J = 4$), ordered from better to worse mental health, as detailed in Table~\ref{tab:ghq-12}.

\begin{table*}[!h]
\begin{center}
\resizebox{16.5cm}{!} {
\begin{tabular}{llc}
\toprule
\textbf{GHQ--12 item} & \textbf{Question wording: \textit{Have you\dots?}} & \textbf{Type of response}  \\
\midrule
1. Concentrate                    & \dots been able to concentrate on what you were doing? & I \\
2. Lose sleep over worries          & \dots lost a lot of sleep because of worries? & II \\
3. Play a useful role             & \dots felt that you have been playing a useful role in life? & I \\
4. Make decisions                 & \dots felt capable of making decisions? & I \\
5. Constantly under strain        & \dots constantly felt overwhelmed and under stress? & II \\
6. Unable to overcome difficulties   & \dots felt unable to overcome your difficulties? & II \\
7. Enjoy activities               & \dots been able to enjoy your daily activities? & I \\
8. Face up to problems            & \dots been able to face up to your problems? & I \\
9. Feel depressed                      & \dots felt unhappy or depressed? & II \\
10. Lose confidence               & \dots lost confidence in yourself? & II \\
11. Feel worthless      & \dots felt that you are a worthless person? & II \\
12. Feel reasonably happy         & \dots felt reasonably happy, considering all the circumstances? & I \\
\bottomrule
\end{tabular}
}
\caption{12-item General Health Questionnaire (GHQ--12) included in the HSRV2022. For positively worded questions (Type I), the response scale follows the pattern ``More / Same / Less / Much less than usual''; whereas for negatively framed questions (Type II), it follows ``Not at all / No more / Rather more / Much more than usual''. Anyway, all categorical variables have response levels ordered from better to worse mental health status.}
\label{tab:ghq-12}
\end{center}
\end{table*}

In Spain, health surveys are conducted periodically at both national and regional levels. Specifically, in the RV, these surveys are carried out approximately every five years, with the most recent available data collected in 2022. The HSRV provides a tool for monitoring the health of this region, collecting information on health behaviors, mental health conditions, physical limitations or social support needs. The sampling design used in the 2022 edition of the survey is described below, as it will be incorporated into the statistical analysis. A total of $n = 9797$ surveys were conducted following a stratified sampling scheme. In particular, about 400 interviews were conducted in each health department (out of a total of 24) among residents aged 15 and older. Independent samples were drawn for the population subgroups defined by age ranges (15--64 and 65+ years) and sex. Additionally, the size of the municipality of residence was taken into account. Specifically, within each sex-age stratum of each health department, the sample was selected through systematic sampling with a random start, applied to the population ordered by municipality size. For our statistical analysis, we will ignore the systematic component of the sampling, assuming therefore simple random sampling within sex-age strata, so the design can be ignored given that those variables are included in the model.

To obtain small-area estimates with substantial geographical disaggregation, we define the municipality as the spatial unit of interest, with a total of $M = 542$ municipalities in the RV. Let $\boldsymbol{x}_i = (s_i, a_i, m_i)^{T}$ be the vector containing the values of the categorical covariates used in the sampling design and the small area (municipality) for the $i$-th respondent, where $s_i \in \{1, 2\}$ denotes the sex: $1$ $=$ Male, $2$ $=$ Female; $a_i \in \{1, \dots, 8\}$ denotes the age group: $1$ $=$ $[15,25)$, $2$ $=$ $[25,35)$, $3$ $=$ $[35,45)$, $4$ $=$ $[45,55)$, $5$ $=$ $[55,65)$, $6$ $=$ $[65,70)$, $7$ $=$ $[70,75)$, $8$ $=$ $[75,\dots)$; and $m_i \in \{1, \dots, 542\}$ denotes the municipality.

From now on, we will consider the three multivariate models described in Sections~\ref{sec:indep}, ~\ref{sec:corr} and ~\ref{sec:corrandire}, respectively. In this case study, spatial contiguity between municipalities defines the adjacency structure used for the corresponding spatial random effects. In all cases, rather than assuming fixed effects for the categorical covariates $s_i$ and $a_i$, we allow them to have different effects on the cumulative probabilities; that is, individuals with different values of sex and age group are assigned distinct, fully independent, cut points, as already suggested in Section~\ref{sec:methods}.

So, following the guidelines in Section~\ref{sec:indep}, the first model (which we will refer to as the Independent Model, or simply Model--Indep) assumes: 
\begin{equation}
\logit (\gamma_{jk}(\boldsymbol{x}_i)) = \kappa_{s_i a_i j k} + \theta_{m_i k},
\label{eq:cs-indep}
\end{equation}
where the collection $\{\kappa_{s a j k}\}_{j = 1}^{J-1}$ represents the average value of the associated cumulative probabilities on the logit scale for each combination of sex $s$, age group $a$ and variable $k$; and $\{\theta_{m k}\}_{m = 1}^{M}$ denotes the collection of spatial random effects for the $k$-th question. The prior distribution for vector $\boldsymbol{\kappa}_{s a k} = (\kappa_{sa1k},\dots,\kappa_{sa,J-1,k})^{T}$ is defined according to the steps outlined in Section~\ref{sec:indep} for each sex, age group and variable. As for the spatial random effects, recall that we assume independence across the spatial random vectors $\boldsymbol{\theta}_{k} = (\theta_{1k}, \dots, \theta_{Mk})^{T}$. 

In the second model (which we will refer to as the Correlated Model, or simply Model--Corr) the cumulative probabilities are modeled according to the specifications in Section~\ref{sec:corr}. In particular:
\begin{equation}
\logit (\gamma_{jk}(\boldsymbol{x}_i)) = \kappa_{s_i a_i j k} + \theta_{m_i k},
\label{eq:cs-corr}
\end{equation} 
where $\{\kappa_{s a j k}\}_{j = 1}^{J-1}$ has the same interpretation as in Model--Indep. However, in this setting, the spatial random vectors $\boldsymbol{\theta}_{k}$ capture not only geographic dependencies across small areas for each health indicator, but also correlations among them.

Finally, in the third model (which we will refer to as the Correlated with IREs Model, or simply Model--Corr\&IRE), the cumulative probabilities are defined according to Section~\ref{sec:corrandire}; that is:
\begin{equation}
\logit (\gamma_{jk}(\boldsymbol{x}_i)) = \kappa_{s_i a_i j k} + \theta_{m_i k} + \psi_{ik},
\label{eq:cs-corr-ire}
\end{equation} 
where $\{\kappa_{s a j k}\}_{j = 1}^{J-1}$ and $\{\theta_{m k}\}_{m = 1}^{M}$ have the same interpretation as in Model--Corr. Here, the individual random vectors $\boldsymbol{\psi}_{k} = (\psi_{1k}, \dots, \psi_{nk})^{T}$ account for respondent variability for each health indicator, as well as any correlations among them.

All models have been implemented in the MCMC package \texttt{NIMBLE}. \cite{Valpine2017} The code used to run the entire study can be found at
\href{https://github.com/bsmiguelangel/a-multivariate-spatial-model-for-ordinal-survey-based-data}{https://github.com/bsmiguelangel/a-multivariate-spatial-model-for-ordinal-survey-based-data} or in the Supplemental material section at the end of this article. For each model, we ran five chains in parallel with $8000$ iterations per chain. Of these, the first $2000$ iterations were discarded as burn-in, and the resulting chains were thinned to retain one out of every $30$ iterations. Therefore, a total of $1000$ simulations ($200$ per chain) from the posterior distribution were saved for each parameter of interest. Convergence was assessed by means of visual checks, the Gelman--Rubin statistic and the effective sample size. Specifically, the Gelman--Rubin statistic was not greater than $1.10$ and the effective number of simulations was not lower than $100$ for any of the identifiable parameters in the models.

\begin{figure}[!h]
%\centering
\hspace{-3cm}
\includegraphics[width = 23cm]{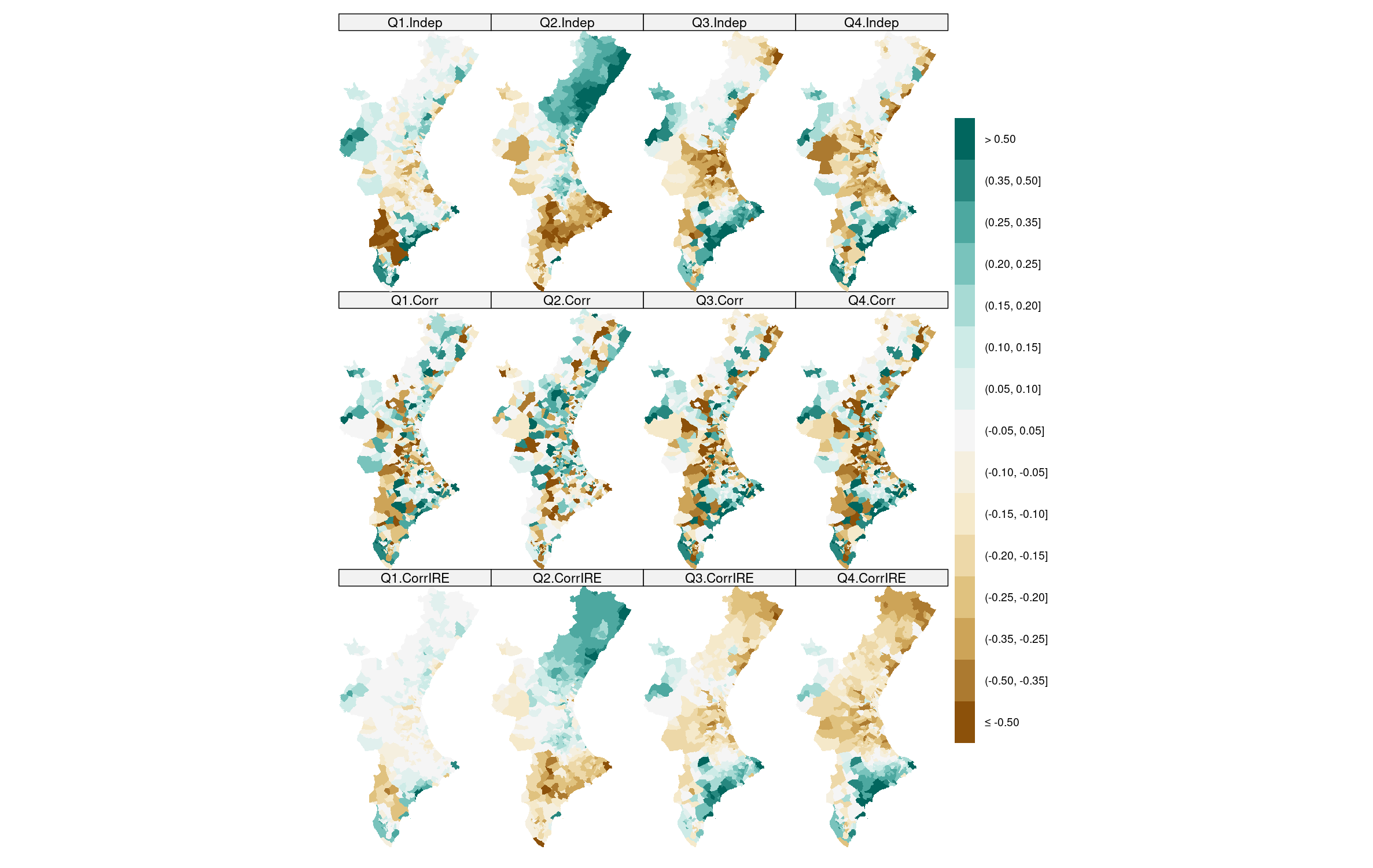}
\vspace{-0.5cm}
\caption{Posterior means of the spatial random vectors $\boldsymbol{\theta}_k$ for $k = 1, \dots, 4$, obtained with the three proposed multivariate models.}
\label{fig:BeltranSanchez1}
\end{figure}

Figure~\ref{fig:BeltranSanchez1} shows the geographic distribution of the spatial random effects $\{\theta_{m k}\}_{m = 1}^M$ for the first four variables in the block; that is, for $k = 1, \dots, 4$. Specifically, we display their posterior means for the three models previously described. This enables the identification of municipalities with particularly worse or better mental health status for each condition. Indeed, green (brown) indicates municipalities with better (worse) mental health status according to each variable. Note that parameters $\theta_{m k}$'s capture the geographic variability not explained by the covariates; that is, they reflect the remaining spatial pattern after controlling for the effect of sex and age group on the available sample. We observe greater spatial smoothness in both Model--Indep and Model--Corr\&IRE compared to Model--Corr, whose geographical patterns appear noticeably noisier. It seems that forcing correlation between variables without accounting for individual-level effects distorts the areal estimates, transferring individual variability to heterogeneity between areas, and therefore resulting in much noisier patterns. As a consequence, it seems important to structure individual and spatial variability separately in order to obtain sensible areal estimates. Moreover, Model--Corr\&IRE exhibits more clearly defined geographical patterns than Model--Indep, as it borrows strength across responses at both the spatial and individual levels. Ultimately, Model--Corr\&IRE improves the estimation of spatial patterns by borrowing strength across all questions within the block, while also demonstrating the importance of including IREs to filter out potential individual-level correlations among responses. Note that the choice of one model over another appears to have a substantial influence on the estimated geographical patterns. The spatial patterns associated with the eight remaining variables of the block can be found in the Supplemental material section at the end of this article.

\begin{figure}[!h]
%\centering
\hspace{-3.5cm}
\includegraphics[width = 23cm]{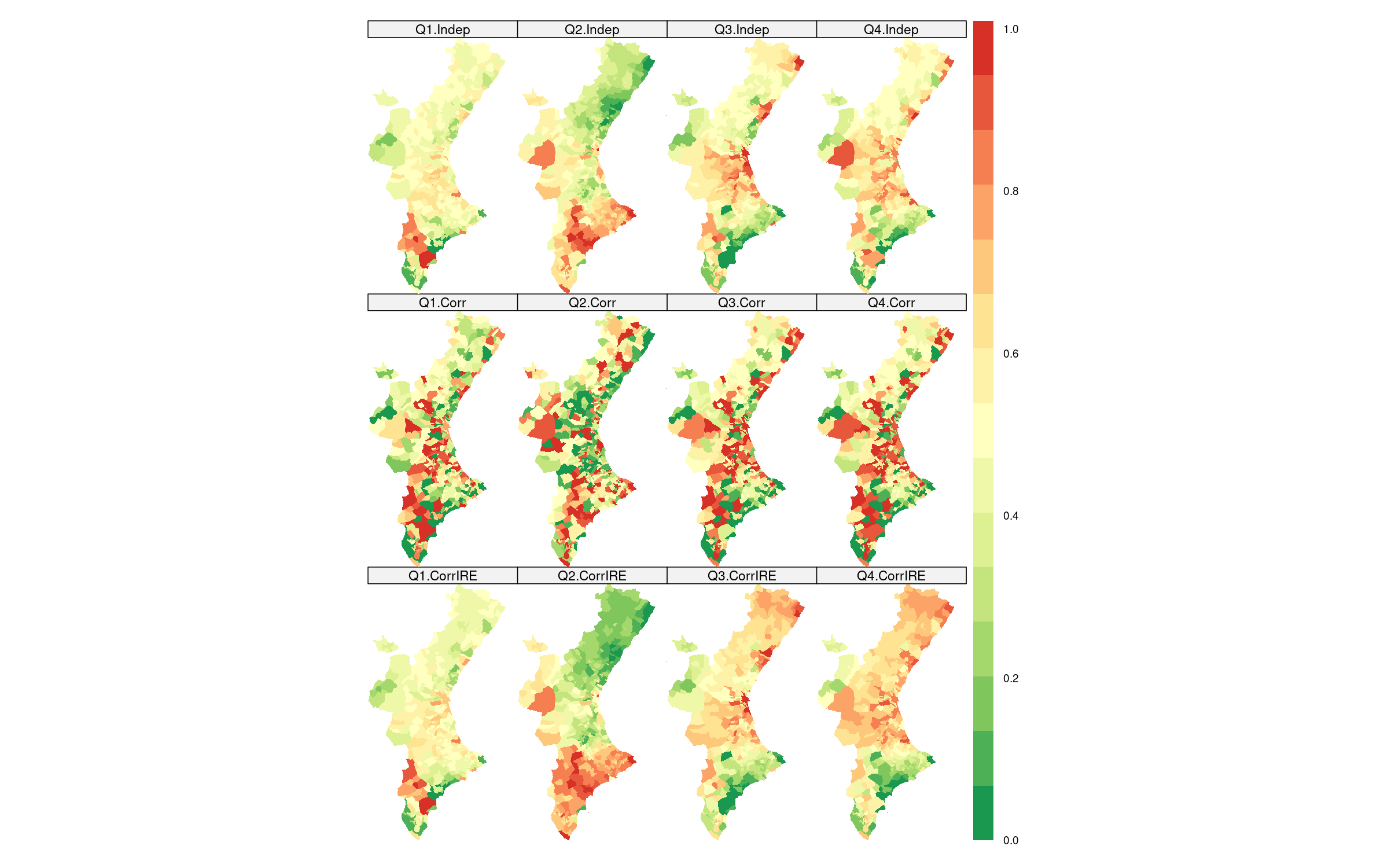}
\vspace{-0.5cm}
\caption{Posterior relevances of the spatial random vectors $\boldsymbol{\theta}_k$ for $k = 1, \dots, 4$, obtained with the three proposed multivariate models.}
\label{fig:BeltranSanchez2}
\end{figure}

Figure~\ref{fig:BeltranSanchez2} displays the relevance of the spatial random effects $\{\theta_{m k}\}_{m = 1}^M$ for the first four variables in the block; that is, for $k = 1, \dots, 4$. Specifically, we measure this relevance as $P(\theta_{m k} < 0|\{\boldsymbol{y}_i\}_{i = 1}^{9797})$ for $m = 1, \dots, 542$ (posterior probability that mental health status is worse than in the region as a whole). Values of that probability close to one (zero) imply a high (low) probability that the associated spatial random effect is negative. Recall that negative values of the $\theta_{m k}$'s make the cut points $\kappa_{s a j k}$'s smaller (lower cumulative probabilities), which increases the probability of the latter categories. In particular, the probability of the first category decreases, while the last one increases. In summary, for those municipalities in Figure~\ref{fig:BeltranSanchez2} coloured in red, we have evidence of worse mental health status, while for those in green we find evidence of better mental health. For example, we could consider that the spatial random effect is relevant if that probability is greater than $0.80$ or less than $0.20$. The spatial smoothness of these relevance patterns closely resembles that of the random effects of Figure~\ref{fig:BeltranSanchez1}. Note that we find stronger evidences in Model--Corr\&IRE than in Model--Indep, as the former borrows strength across responses at both the spatial and individual levels. Further details on the relevance patterns for the remaining variables of the block are provided in the Supplemental material at the end of this article.

\begin{figure}[!h]
%\centering
\hspace{-1cm}
\includegraphics[width = 18cm]{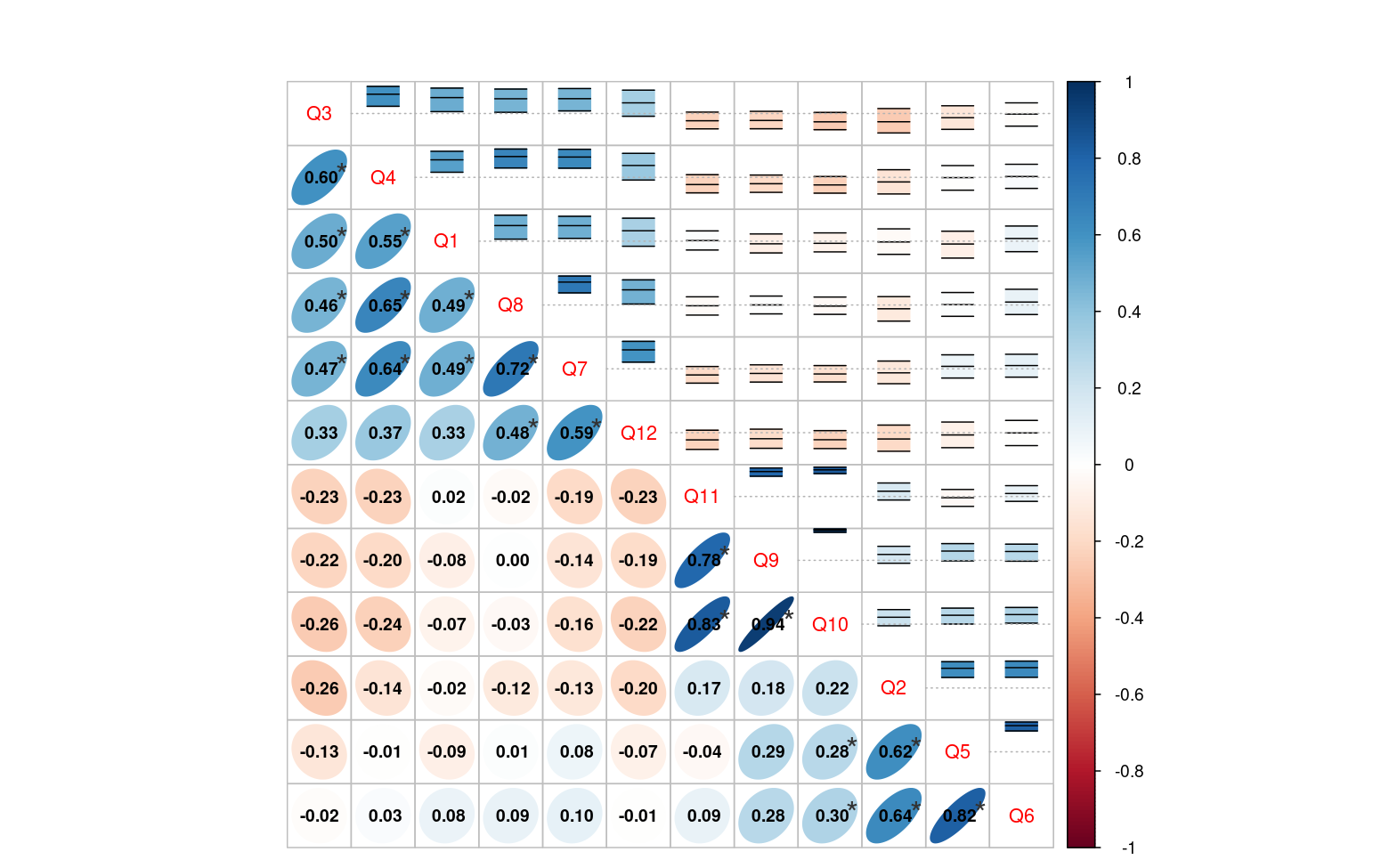}
\vspace{-0.5cm}
\caption{Correlations among responses at the municipality level under Model--Corr\&IRE. Lower triangular shows posterior means of those correlations, while the upper triangular displays their corresponding $95\%$ posterior credible intervals.}
\label{fig:BeltranSanchez3}
\end{figure}

Figure~\ref{fig:BeltranSanchez3} displays the municipality-level correlations $Cor(\boldsymbol{\theta}_k,\boldsymbol{\theta}_{k'})$ for each pair of mental health indicators under Model--Corr\&IRE. These correlations are computed from the variance-covariance matrix $\boldsymbol{\Sigma}_{b} = \boldsymbol{M}^T \boldsymbol{M}$. The lower triangular part of the matrix shows the posterior means of the correlations. An asterisk is included when a correlation is statistically relevant (the $95\%$ credible interval does not include zero). The upper triangular part displays boxes whose extremes are the $0.025$ and $0.975$ quantiles of the posterior samples of the corresponding correlations. If the boxes contain the dashed line at zero, the correlation is not considered relevant. We have ordered here the twelve questions using an agglomerative method, specifically Ward's clustering algorithm. \cite{Murtagh2014} This procedure yielded two clusters, each comprising six responses that are positively correlated at the municipality level. On the one hand, items 1, 3, 4, 7, 8, and 12 from Table~\ref{tab:ghq-12} form a sub-block commonly referred to in the literature as ``social dysfunction''\cite{Politi1994, Doi2003,  Gao2012, Glozah2015}, ``loss of positive emotions''\cite{Suzuki2011} or ``general functioning''\cite{Centofanti2019}. On the other hand, items 2, 5, 6, 9, 10, and 11 from Table~\ref{tab:ghq-12} correspond to a sub-block known as ``general dysphoria''\cite{Politi1994, Centofanti2019}, ``psychological distress''\cite{Doi2003, Gao2012} or ``depression/anxiety''\cite{Suzuki2011, Glozah2015}. Ultimately, this suggests that in areas where mental health issues related to one item of the sub-block are present, problems in other items within the same sub-block are also likely to occur. Our model was able to detect the two sub-blocks of questions, repeatedly referenced in the literature related to this questionnaire, and to take advantage of the dependencies within each block, but not across them.

\begin{figure}[!h]
%\centering
\hspace{-1cm}
\includegraphics[width = 18cm]{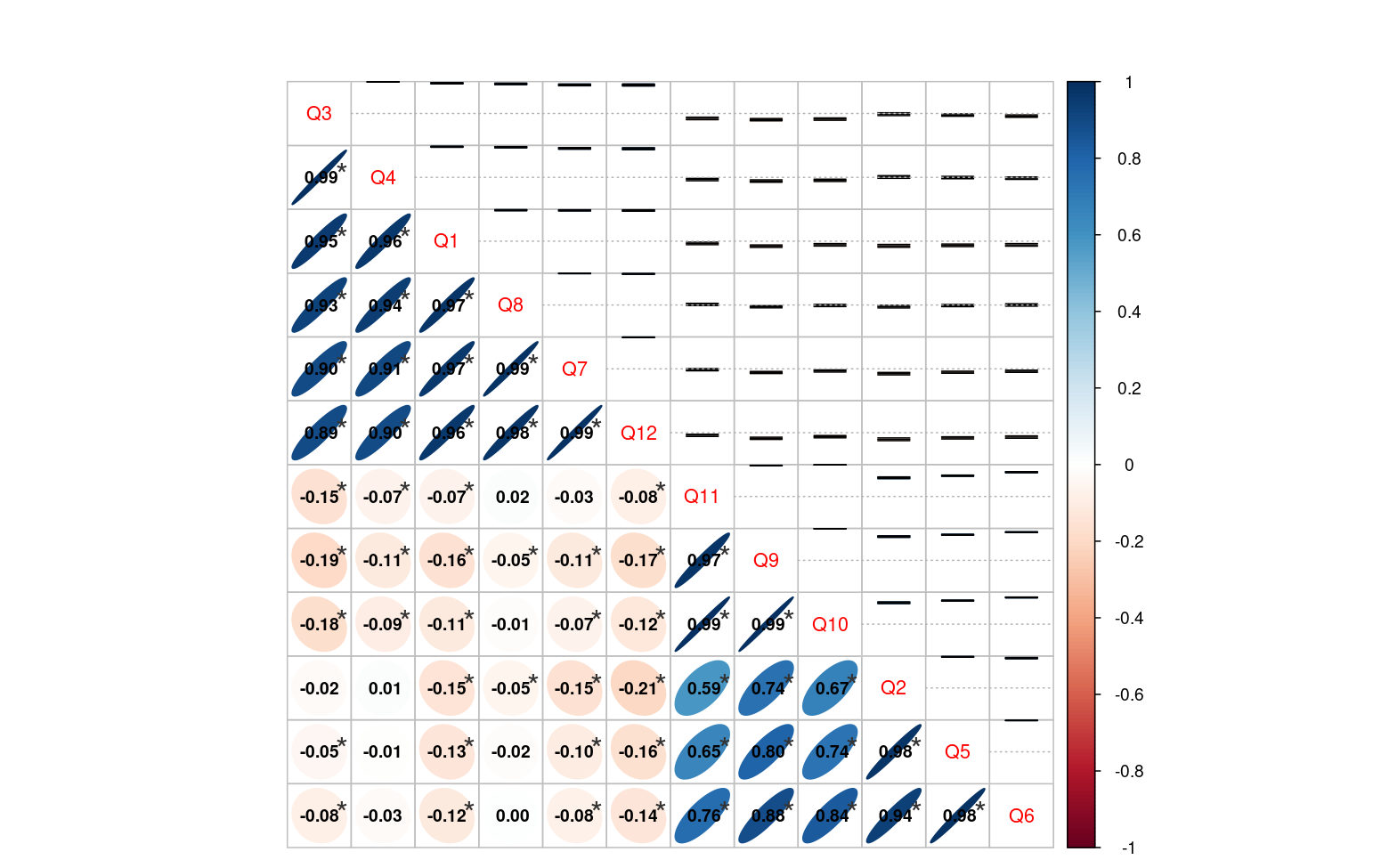}
\vspace{-0.5cm}
\caption{Correlations among responses at the individual level under Model--Corr\&IRE. Lower triangular shows posterior means of those correlations, while the upper triangular displays their corresponding $95\%$ posterior credible intervals.}
\label{fig:BeltranSanchez4}
\end{figure}

Figure~\ref{fig:BeltranSanchez4} shows the individual-level correlations $Cor(\boldsymbol{\psi}_k,\boldsymbol{\psi}_{k'})$ for each pair of variables in the mental health block corresponding to Model--Corr\&IRE. These correlations are derived from the individual-level variance-covariance matrix $\boldsymbol{\tilde{\Sigma}}_{b} = \boldsymbol{\tilde{M}}^T \boldsymbol{\tilde{M}}$. The sub-blocks identified at the small-area level also emerge at the individual level. However, the strength of association is noticeably higher in this case. This indicates a tendency for individuals with mental health problems in one question of a sub-block to also report problems with related items. It is also worth mentioning that including IREs allows us to account for (and remove) potential correlations between responses that are due exclusively to the respondents' variability, thereby enabling more accurate estimation of correlations at the small-area level. In fact, the small-area correlation matrix in Model--Corr (see Supplemental material) closely resembles that in Figure~\ref{fig:BeltranSanchez4}, clearly illustrating the distorting effect of individual variability on the spatial effect correlations, an issue that Model--Corr\&IRE is able to fix.

\begin{figure}[!h]
\centering
\includegraphics[width = 12cm]{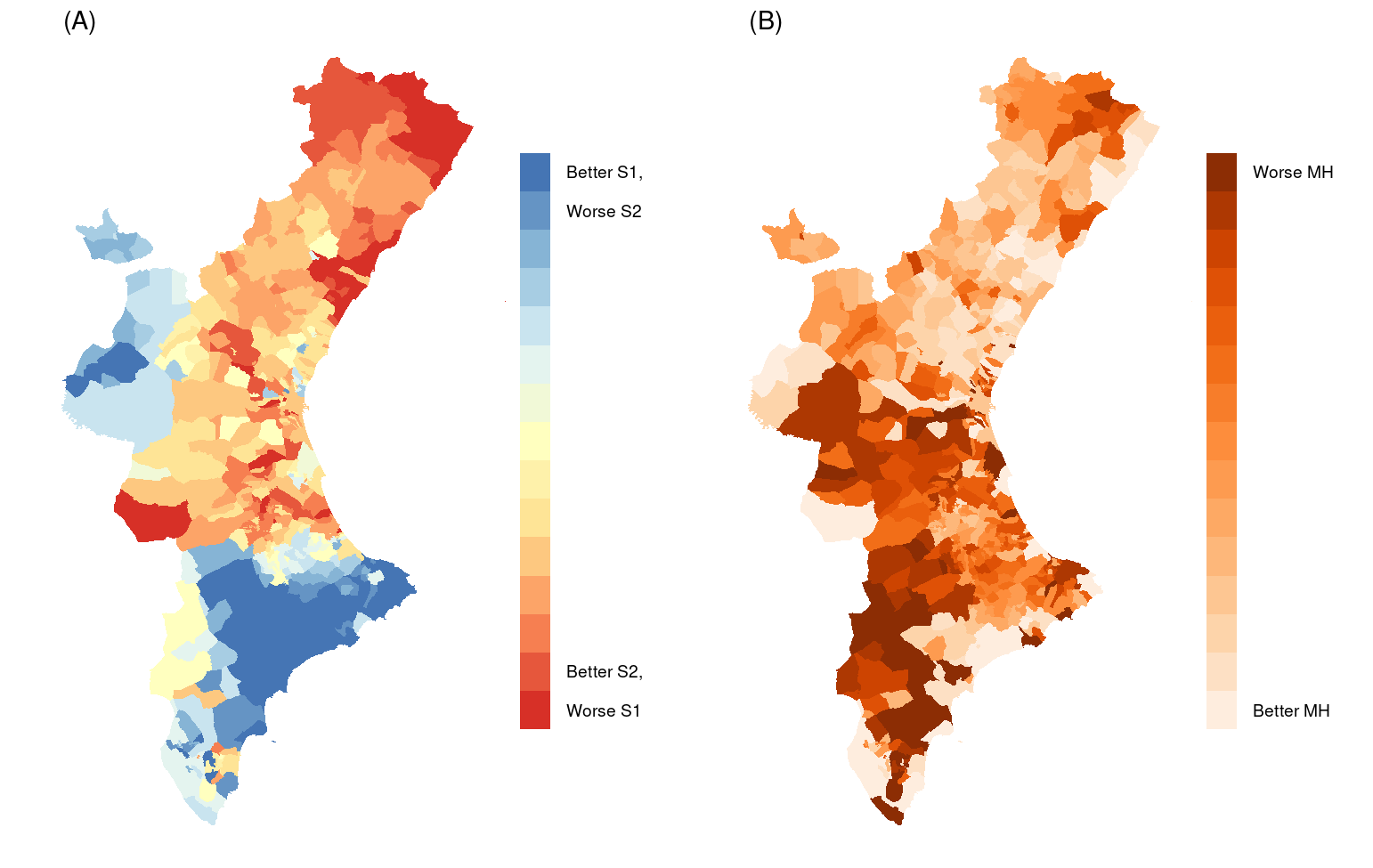}
\vspace{-0.5cm}
\caption{(A) First and (B) second principal components from a principal component analysis (PCA) of the posterior means of the spatial random effects $\theta_{m k}$'s under Model--Corr\&IRE. ``S1'' (sub-block $1$) includes GHQ--12 items 1, 3, 4, 7, 8, and 12 from Table~\ref{tab:ghq-12}, while ``S2'' (sub-block $2$) comprises items 2, 5, 6, 9, 10, and 11. ``MH'' simply stands for mental health in general. Thus, for example, ``Better S1, Worse S2'' should be interpreted as better mental health status in questions from S1, and/or worse mental health status in those from S2.}
\label{fig:BeltranSanchez5}
\end{figure}

In addition to Figure~\ref{fig:BeltranSanchez1}, where we have represented the geographical patterns corresponding to four variables of the block, it may also be useful to visualize the principal components from a principal component analysis (PCA) of the municipality-level posterior means across all spatial patterns. According to the results shown in the correlation matrix of Figure~\ref{fig:BeltranSanchez3}, two principal components would appropriately summarize the information from the twelve original variables of the block. Based on Model--Corr\&IRE, Figure~\ref{fig:BeltranSanchez5} (A) displays the geographical pattern for the first principal component derived from the posterior mean of the matrix $\boldsymbol{\Theta}$, allowing us to distinguish regions with better or worse mental health in terms of the two sub-blocks. This map highlights regions that exhibit good mental health for the first sub-block of indicators and/or poor mental health for the second, versus those with the opposite pattern. It should be noted that the first principal component captures shape rather than size, which is unusual. Figure~\ref{fig:BeltranSanchez5} (B) shows the geographical pattern of the second principal component, which provides a map that distinguishes regions with overall better or worse mental health. Unlike the first, this component reflects size and identifies regions with worse mental health across all indicators on average. However, the contrast between sub-blocks appears more informative than the shared pattern.

Furthermore, we evaluate the fit of the three proposed multivariate models (Indep, Corr and Corr\&IRE) by simulating from the posterior predictive distribution for each of the twelve ordinal variables and each respondent, and aggregating the predicted values at the municipality level. Once a particular model has been fitted, we obtain simulations of the posterior probabilities for each respondent from the three models: $\{\hat{\pi}_{jk}^{(r)}(\boldsymbol{x}_i)\}_{r = 1}^{1000}$, $j = 1, \dots, 4$, $k = 1, \dots, 12$, and $i = 1, \dots, 9797$. Subsequently, we simulate values of the corresponding (categorical) response variable from the sampled probabilities. Finally, the number of respondents sampled in each category is computed by aggregating the previously generated values for each municipality. Those frequencies are compared with the (real) observed counts in the sample in order to assess the fit of the models. For example, Table~\ref{tab:assessment} contains, for the $5$-th item of the GHQ--12, the corresponding posterior means, $95\%$ prediction intervals and the observed sample percentages for the six municipalities with the largest sample sizes in the HSRV2022. As observed, there are no large discrepancies between the observed and predicted values. Combinations where the $95\%$ prediction interval excludes the observed sample value are highlighted in red. Note that the prediction intervals are considerably narrower with Model--Corr\&IRE, and they tend to align more closely with the observed values. More information on the model fit for the remaining variables can be found in the Supplemental material section at the end of this article. 

Finally, we have computed the Widely Applicable Information Criterion (WAIC)\cite{Watanabe2010} for the three models. The WAIC values are $192525.4$ for Model--Indep, $190447.6$ for Model--Corr, and $103170.1$ for Model--Corr\&IRE. All of these results indicate a substantial improvement in model fit when both spatial correlation and IREs are incorporated.

\begin{table}[!h]
\begin{center}
\resizebox{16.5cm}{!} {
\begin{tabular}{llllll}
\toprule
\multirow{2}{*}{\textbf{Municipality}} & \multirow{2}{*}{\textbf{Model}} & \multicolumn{4}{c}{\textbf{Response levels}} \\
\cmidrule(lr){3-6}
& & \textbf{Not at all} & \textbf{No more than usual} & \textbf{Rather more than usual} & \textbf{Much more than usual} \\

\midrule
\multirow{4}{*}{Valencia} 
& Indep       & 17.81 (15.18--20.55) & 56.02 (52.91--59.27) & 20.73 (18.00--23.73) & 5.43 (4.00--7.00) \\
& Corr       & \cellcolor{gray!20}17.18 (14.45--19.91) & 56.58 (53.45--59.82) & 20.91 (18.09--23.64) & 5.32 (4.00--6.82) \\
& Corr\&IRE & 17.41 (15.73--19.09) & \cellcolor{gray!20}57.31 (54.82--59.64) & \cellcolor{gray!20}20.50 (18.45--22.55) & \cellcolor{gray!20}4.78 (3.91--5.73) \\
&             & \textbf{16.91}       & \textbf{58.55}       & \textbf{20.18}       & \textbf{4.27} \\
\midrule
\multirow{4}{*}{Alicante} 
& Indep       & 15.44 (12.03--18.81) & 54.87 (50.34--59.15) & 23.25 (19.49--27.29) & \textcolor{red}{6.43 (4.41--8.81)} \\
& Corr       & 15.36 (12.37--18.64) & 55.48 (51.36--59.83) & 23.02 (19.15--26.61) & \textcolor{red}{6.14 (4.24--8.31)} \\
& Corr\&IRE & \cellcolor{gray!20}16.66 (14.41--18.98) & \cellcolor{gray!20}53.71 (50.68--56.95) & \cellcolor{gray!20}21.28 (18.47--24.07) & \cellcolor{gray!20}8.36 (6.95--9.83) \\
&             & \textbf{16.95}       & \textbf{53.39}       & \textbf{20.51}       & \textbf{9.15} \\
\midrule
\multirow{4}{*}{Elche} 
& Indep       & 14.83 (11.55--18.28) & \textcolor{red}{54.63 (50.51--58.79)} & 23.85 (20.00--28.11) & \textcolor{red}{6.68 (4.48--8.97)} \\
& Corr       & 14.57 (11.55--18.10) & \textcolor{red}{55.15 (50.86--59.48)} & 23.84 (19.82--27.76) & \textcolor{red}{6.44 (4.48--8.62)} \\
& Corr\&IRE & \cellcolor{gray!20}16.62 (14.48--18.79) & \cellcolor{gray!20}51.35 (48.28--54.31) & \cellcolor{gray!20}23.47 (20.69--26.03) & \cellcolor{gray!20}8.57 (7.07--10.17) \\
&             & \textbf{17.93}       & \textbf{50.00}       & \textbf{22.07}       & \textbf{9.48} \\
\midrule
\multirow{4}{*}{\makecell{Castellón de\\la Plana}}
& Indep       & 21.55 (15.77--28.08) & 56.64 (50.77--62.69) & 17.48 (12.31--23.09) & 4.33 (1.92--7.31) \\
& Corr       & 21.38 (15.77--27.31) & 57.23 (51.54--63.46) & 17.29 (12.31--22.69) & 4.10 (1.92--6.92) \\
& Corr\&IRE & \cellcolor{gray!20}20.02 (16.15--23.85) & \cellcolor{gray!20}60.72 (55.77--65.38) & \cellcolor{gray!20}16.68 (13.08--20.77) & \cellcolor{gray!20}2.58 (1.54--4.23) \\
&             & \textbf{20.00}       & \textbf{61.54}       & \textbf{16.15}       & \textbf{2.31} \\
\midrule
\multirow{4}{*}{Orihuela} 
& Indep       & 15.88 (10.42--21.89) & 55.09 (48.44--61.98) & 22.75 (16.15--29.69) & \textcolor{red}{6.28 (3.12--10.42)} \\
& Corr       & 17.37 (11.98--23.96) & 56.96 (50.63--65.55) & 20.50 (14.57--26.56) & 5.17 (2.08--8.85) \\
& Corr\&IRE & \cellcolor{gray!20}19.66 (16.15--23.44) & \cellcolor{gray!20}51.82 (46.88--56.77) & \cellcolor{gray!20}23.82 (19.27--28.12) & \cellcolor{gray!20}4.70 (2.60--7.29) \\
&             & \textbf{18.23}       & \textbf{52.08}       & \textbf{26.56}       & \textbf{2.60} \\
\midrule
\multirow{4}{*}{Torrevieja} 
& Indep       & 15.30 (9.78--21.21)  & 54.48 (47.28--61.96) & 23.56 (16.85--30.45) & 6.65 (3.25--10.87) \\
& Corr       & 15.49 (10.31--21.74) & 55.40 (47.83--63.04) & 23.03 (16.29--30.43) & 6.09 (2.72--10.33) \\
& Corr\&IRE & \cellcolor{gray!20}19.88 (16.30--23.37) & \cellcolor{gray!20}48.25 (43.48--53.80) & \cellcolor{gray!20}21.89 (17.39--26.63) & \cellcolor{gray!20}9.98 (7.61--12.50) \\
&             & \textbf{19.02}       & \textbf{49.46}       & \textbf{21.20}       & \textbf{10.33} \\
\bottomrule
\end{tabular}
}
\vspace{0.25cm}
\caption{Model assessment for item 5 of the GHQ--12. Posterior means ($95\%$ prediction intervals) and observed values (in \textbf{bold}) of the percentage of respondents in each category across the selected municipalities in the RV. The model shaded in gray represents the best fit, as its prediction interval includes the observed value and its posterior mean is the closest among all models. Models whose prediction interval does not include the observed value are highlighted in red.}
\label{tab:assessment}
\end{center}
\end{table}

\section{Discussion} \label{sec:discussion}

In this article, we have introduced a multivariate extension of a recent Bayesian individual-level methodology for analyzing spatial ordinal data from health surveys. \cite{BeltranSanchez2024} Our proposal represents an advance in the spatial analysis of health data, since it allows for the joint analysis of multiple ordinal outcomes. The main contribution of our proposal is that it accounts not only for spatial correlation among geographic areas, but also for dependencies among multiple ordinal variables within specific thematic blocks. In fact, the proposed methodology enables the simultaneous estimation of the spatial patterns associated with the different health indicators, as well as the relationships among them at both the small-area and individual levels by following the \textbf{M}--model specifications. \cite{BotellaRocamora2015} 

As evidenced, the multivariate approach yields substantial benefits in practical terms, providing considerably improved maps compared to the univariate approach. However, our results also show the importance of model selection in order to obtain sensible results. A wrong structuring of variability sources, such as ignoring individual variability, may easily yield inappropriate geographic estimates, even worse than those obtained from univariate studies. In our opinion, this is an important insight of this paper. Beyond the potential benefits in terms of fit and the evident improvement in the estimation of risk patterns, our paper highlights the risks of conducting poorly specified multivariate analyses. These risks should be carefully considered in future statistical analyses in this context. 

Through the construction of synthetic indicators, such as maps and correlation matrices, that reveal the underlying structure in the data that might otherwise remain hidden under univariate approaches, this new proposal facilitates a more comprehensive use of survey data beyond their original purpose, enhancing their practical utility. In particular, it offers a concise and high-level summary of the detailed, extensive, and often hard-to-interpret information commonly found in health survey reports. We believe this methodology represents a powerful tool to maximize the value of survey data, particularly health surveys, which are routinely conducted worldwide to monitor the health status of large population groups. It is worth mentioning that the proposed methodology could be easily and directly extended to accommodate variables within a thematic block that have different numbers of categories, including binary (Bernoulli) responses. 

Note that we have focused our discussion on the spatial random effects $\theta_{m k}$'s; that is, the remaining geographical patterns after controlling for the effect of covariates on the available sample. If the goal were to estimate finite population quantities (e.g., the proportion of individuals in each health status category at the small-area level), a post-stratification estimator should be used, as proposed in the univariate ordinal framework. \cite{BeltranSanchez2024}

As future lines of research, it would be interesting to consider spatio-temporal extensions of the current proposal to analyze successive editions of a health survey commonly conducted within the same study region. Since it is common for health surveys to be collected periodically (e.g., every five years), such an extension could help identify meaningful temporal trends in population health. Furthermore, it would also be interesting to apply this methodology to national or European health surveys at small-area levels, although doing so would pose significant challenges due to the expected high computational cost. Finally, one limitation of the proposed methodology is that it assumes all design variables are known, which is not always the case in practice. In such situations, the modeling framework would need to be adapted to incorporate survey weights; otherwise, failing to account for the full sampling design information may lead to biased estimates.

\section*{Acknowledgements}

The authors acknowledge the funding of Ministerio de Ciencia, Innovación y Universidades of Spain (MCIN/AEI/10.13039/501100011033/ FEDER, UE) and the European Regional Development Fund.

\section*{Declaration of conflicting interests}

The author(s) declared no potential conflicts of interest with respect to the research, authorship, and/or publication of this article.

\section*{Funding}

This paper is part of the project PID2022-136455NB-I00, funded by Ministerio de Ciencia, Innovación y Universidades of Spain (MCIN/AEI/10.13039/501100011033/ FEDER, UE) and the European Regional Development Fund.

\section*{Data availability}

All code used for the statistical analysis is available as a fully reproducible script at
\href{https://github.com/bsmiguelangel/a-multivariate-spatial-model-for-ordinal-survey-based-data}{https://github.com/bsmiguelangel/a-multivariate-spatial-model-for-ordinal-survey-based-data}. The individual-level nature of the dataset prevents us from making it publicly available. However, access to the data may be granted, upon request, by the data providers under the same conditions as those applied to the authors of this study.

\section*{ORCID iDs}
\noindent
Beltrán-Sánchez, MA \orcidlink{0000-0001-9450-2973} \href{https://orcid.org/0000-0001-9450-2973}{https://orcid.org/0000-0001-9450-2973} \\ 
Martínez-Beneito, MA \orcidlink{0000-0001-8406-8050} \href{https://orcid.org/0000-0001-8406-8050}{https://orcid.org/0000-0001-8406-8050} \\
Corberán-Vallet, A \orcidlink{0000-0002-1091-9534} \href{https://orcid.org/0000-0002-1091-9534}{https://orcid.org/0000-0002-1091-9534}

\newpage

\section*{Supplemental material}

\setcounter{figure}{0}
\renewcommand{\thefigure}{S\arabic{figure}}

\begin{figure}[!h]
\hspace{-1.75cm}
\includegraphics[width = 20cm]{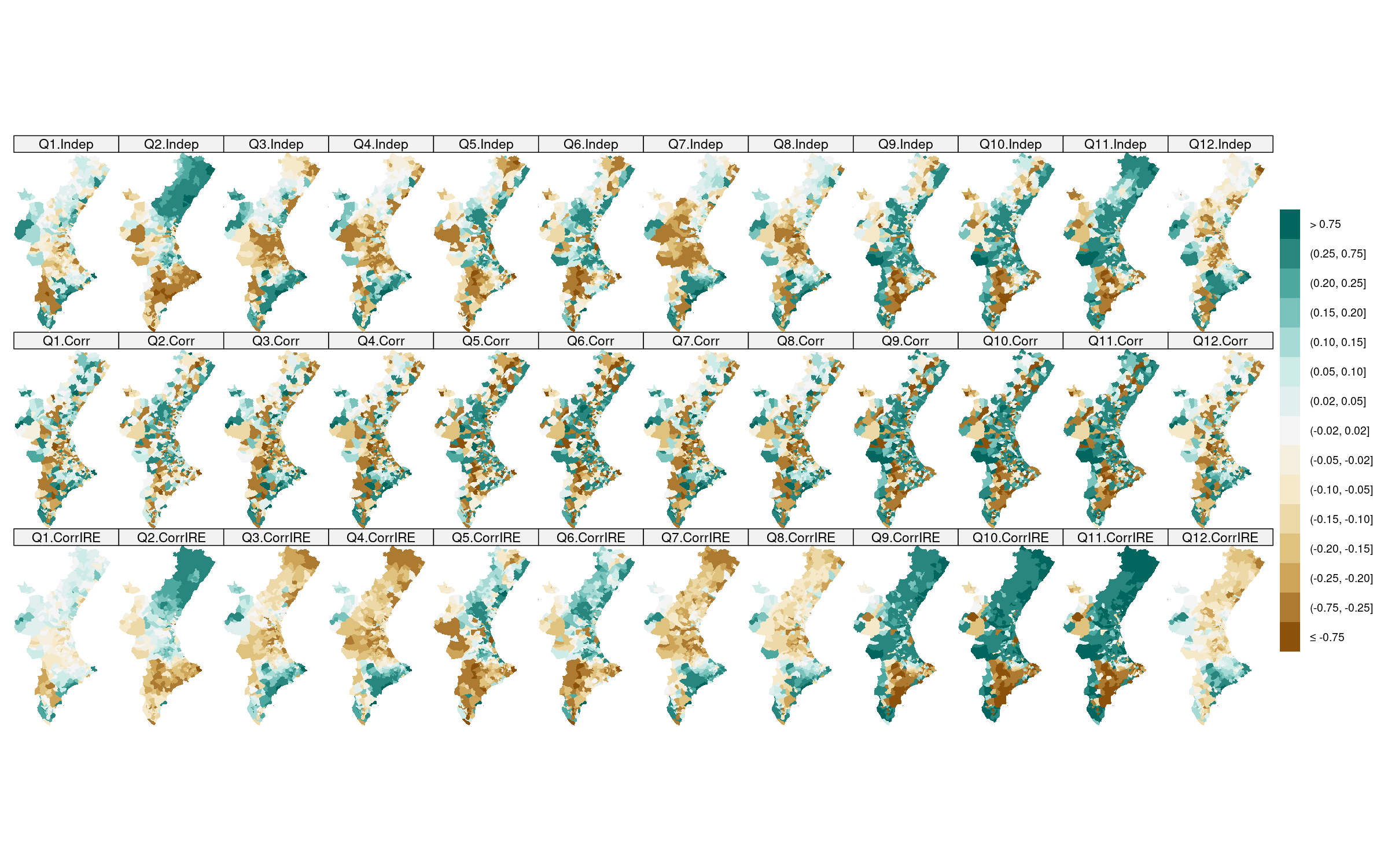}
\vspace{-2.5cm}
\caption{Posterior means of the spatial random vectors $\boldsymbol{\theta}_k$ for $k = 1, \dots, 12$, obtained with the three proposed multivariate models.}
\label{fig:SupplementalMaterial1}
\end{figure}

\begin{figure}[!h]
\hspace{-1.75cm}
\includegraphics[width = 20cm]{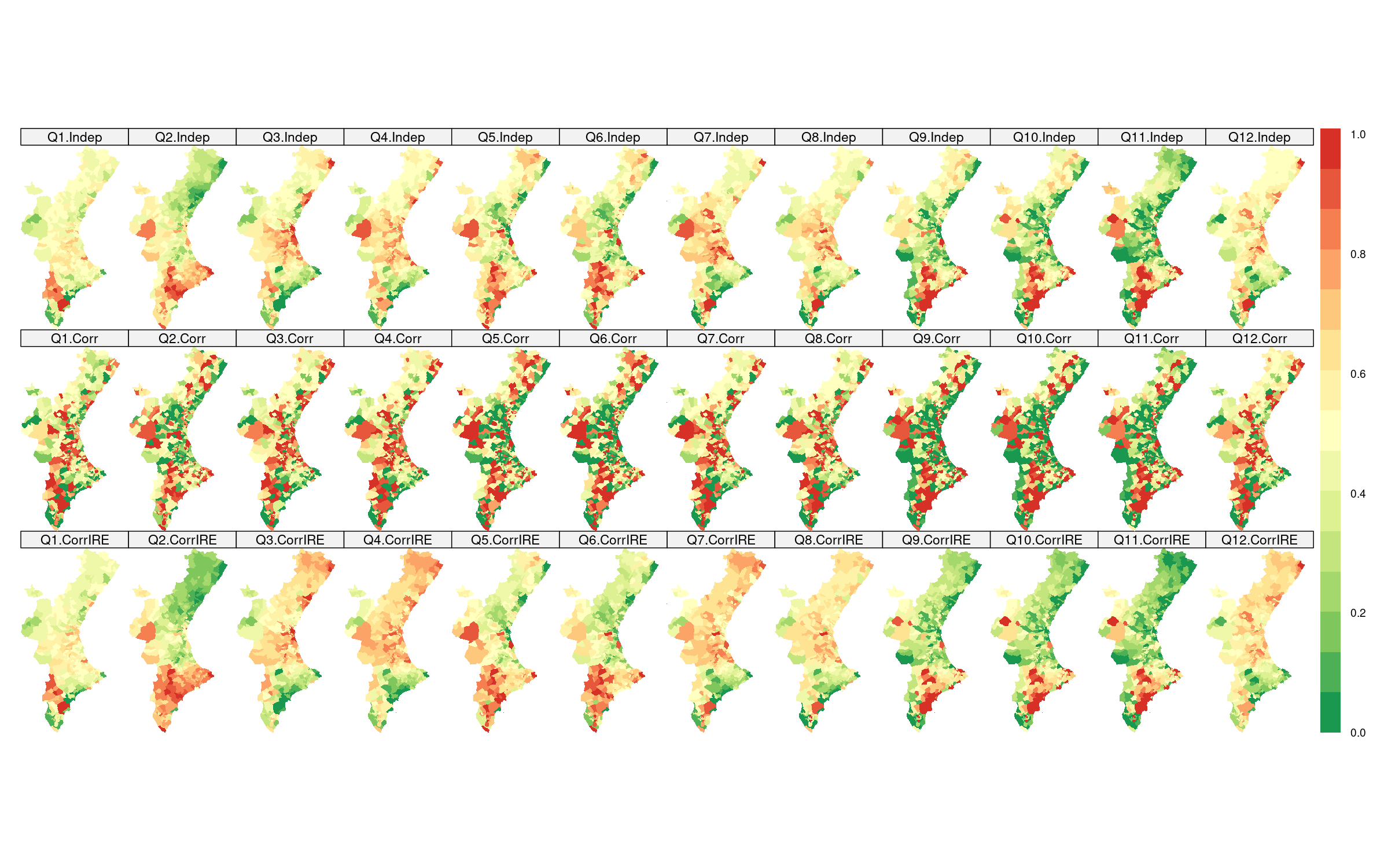}
\vspace{-2.5cm}
\caption{Posterior relevances of the spatial random vectors $\boldsymbol{\theta}_k$ for $k = 1, \dots, 12$, obtained with the three proposed multivariate models.}
\label{fig:SupplementalMaterial2}
\end{figure}

\begin{figure}[!h]
%\centering
\hspace{-1cm}
\includegraphics[width = 18cm]{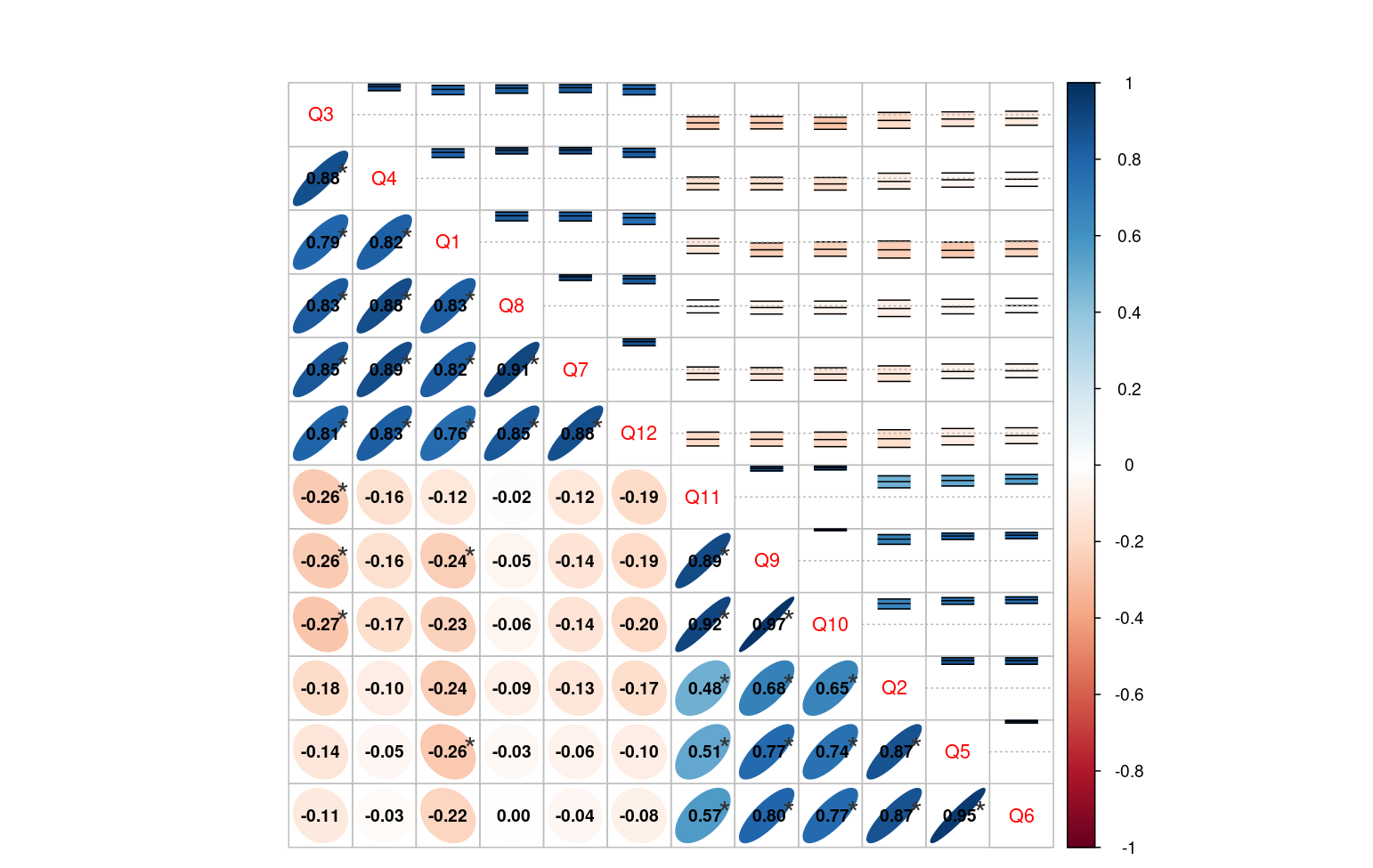}
\vspace{-0.5cm}
\caption{Correlations among responses at the municipality level under Model--Corr. Lower triangular shows posterior means of those correlations, while the upper triangular displays their corresponding $95\%$ posterior credible intervals.}
\label{fig:SupplementalMaterial3}
\end{figure}

\setcounter{table}{0}
\renewcommand{\thetable}{S\arabic{table}}

\begin{table}[!h]
\begin{center}
\resizebox{16.5cm}{!} {
\begin{tabular}{llllll}
\toprule
\multirow{2}{*}{\textbf{Municipality}} & \multirow{2}{*}{\textbf{Model}} & \multicolumn{4}{c}{\textbf{Response levels}} \\
\cmidrule(lr){3-6}
& & \textbf{More than usual} & \textbf{Same as usual} & \textbf{Less than usual} & \textbf{Much less than usual} \\
\midrule

\multirow{4}{*}{Valencia} 
& Indep       & 4.79 (3.36--6.36) & 83.51 (81.09--85.91) & \cellcolor{gray!20}9.54 (7.64--11.64) & 2.17 (1.36--3.18) \\
& Corr        & \cellcolor{gray!20}4.51 (3.27--5.91) & \cellcolor{gray!20}83.74 (81.45--86.00) & 9.61 (7.82--11.64) & 2.14 (1.27--3.09) \\
& Corr\&IRE   & 4.53 (3.36--5.73) & 83.50 (81.18--85.55) & 9.95 (8.18--11.91) & \cellcolor{gray!20}2.02 (1.27--2.91) \\
&            & \textbf{4.36} & \textbf{84.55} & \textbf{9.18} & \textbf{1.91} \\
\midrule

\multirow{4}{*}{Alicante} 
& Indep       & 4.05 (2.37--5.93) & 82.26 (78.81--85.76) & 11.12 (8.14--14.41) & 2.57 (1.35--4.07) \\
& Corr        & \cellcolor{gray!20}3.85 (2.37--5.76) & 82.45 (78.98--85.76) & 11.22 (8.31--14.24) & 2.49 (1.19--3.90) \\
& Corr\&IRE   & 4.23 (2.71--5.76) & \cellcolor{gray!20}81.57 (78.64--84.41) & \cellcolor{gray!20}11.20 (8.64--13.90) & \cellcolor{gray!20}3.00 (1.86--4.41) \\
&            & \textbf{3.90} & \textbf{81.02} & \textbf{12.20} & \textbf{2.88} \\
\midrule

\multirow{4}{*}{Elche} 
& Indep       & \textcolor{red}{5.60 (3.62--7.76)} & \textcolor{red}{84.13 (80.69--87.07)} & 8.36 (5.86--10.86) & 1.91 (0.86--3.28) \\
& Corr        & \textcolor{red}{5.97 (3.97--8.45)} & \textcolor{red}{85.00 (81.90--87.94)} & \textcolor{red}{7.42 (5.17--9.83)} & \textcolor{red}{1.60 (0.69--2.76)} \\
& Corr\&IRE   & \textcolor{red}{7.69 (5.69--9.66)} & \textcolor{red}{81.42 (78.45--84.31)} & \cellcolor{gray!20}8.95 (6.55--11.38) & \cellcolor{gray!20}1.94 (1.03--2.93) \\
&            & \textbf{10.00} & \textbf{76.38} & \textbf{10.17} & \textbf{2.93} \\
\midrule

\multirow{4}{*}{\makecell{Castellón de\\la Plana}} 
& Indep       & 5.27 (2.68--8.46) & 84.03 (78.85--88.47) & \cellcolor{gray!20}8.75 (5.00--13.08) & 1.96 (0.38--3.85) \\
& Corr        & \cellcolor{gray!20}5.13 (2.69--8.09) & 84.26 (79.62--88.85) & 8.73 (4.99--12.70) & 1.88 (0.38--3.47) \\
& Corr\&IRE   & 4.22 (1.92--6.92) & \cellcolor{gray!20}86.26 (81.92--90.00) & 8.09 (4.99--11.54) & \cellcolor{gray!20}1.43 (0.38--2.70) \\
&            & \textbf{5.00} & \textbf{85.77} & \textbf{8.85} & \textbf{0.38} \\
\midrule

\multirow{4}{*}{Orihuela} 
& Indep       & 5.55 (2.60--8.85) & 84.29 (79.17--89.06) & 8.29 (4.17--12.50) & 1.87 (0.00--4.17) \\
& Corr        & 6.64 (3.12--10.42) & 85.40 (80.21--90.10) & \cellcolor{gray!20}6.61 (3.12--10.42) & \cellcolor{gray!20}1.35 (0.00--3.12) \\
& Corr\&IRE   & \cellcolor{gray!20}7.82 (4.69--11.46) & \cellcolor{gray!20}82.41 (77.60--86.98) & 8.18 (4.67--11.98) & 1.59 (0.52--3.12) \\
&            & \textbf{7.81} & \textbf{83.85} & \textbf{6.77} & \textbf{1.04} \\
\midrule

\multirow{4}{*}{Torrevieja} 
& Indep       & 5.86 (2.72--9.78) & 83.97 (78.26--89.13) & 8.34 (4.35--12.50) & 1.83 (0.00--3.80) \\
& Corr        & 6.33 (2.72--10.33) & 84.76 (79.35--90.22) & 7.30 (3.26--11.96) & \cellcolor{gray!20}1.60 (0.00--3.80) \\
& Corr\&IRE   & \cellcolor{gray!20}7.46 (4.35--10.87) & \cellcolor{gray!20}82.21 (77.17--86.97) & \cellcolor{gray!20}8.70 (4.89--13.04) & 1.62 (0.54--3.26) \\
&            & \textbf{8.15} & \textbf{82.07} & \textbf{8.70} & \textbf{1.09} \\
\bottomrule
\end{tabular}
}
\vspace{0.25cm}
\caption{Model assessment for item 1 of the GHQ--12. Posterior means ($95\%$ prediction intervals) and observed values (in \textbf{bold}) of the percentage of respondents in each category across the selected municipalities in the RV. The model shaded in gray represents the best fit, as its prediction interval includes the observed value and its posterior mean is the closest among all models. Models whose prediction interval does not include the observed value are highlighted in red.}
\label{tab:assessment-1}
\end{center}
\end{table}

\begin{table}[!h]
\begin{center}
\resizebox{16.5cm}{!} {
\begin{tabular}{llllll}
\toprule
\multirow{2}{*}{\textbf{Municipality}} & \multirow{2}{*}{\textbf{Model}} & \multicolumn{4}{c}{\textbf{Response levels}} \\
\cmidrule(lr){3-6}
& & \textbf{Not at all} & \textbf{No more than usual} & \textbf{Rather more than usual} & \textbf{Much more than usual} \\
\midrule

\multirow{4}{*}{Valencia} 
& Indep       & \cellcolor{gray!20}16.93 (14.36--19.73) & 66.47 (63.45--69.45) & 11.43 (9.36--13.64) & 5.17 (3.82--6.73) \\
& Corr        & 17.18 (14.36--19.91) & \cellcolor{gray!20}66.92 (64.00--69.82) & \cellcolor{gray!20}11.05 (9.18--13.18) & 4.85 (3.54--6.27) \\
& Corr\&IRE   & 17.02 (14.91--19.18) & 66.46 (63.55--69.09) & 11.73 (9.82--13.82) & \cellcolor{gray!20}4.79 (3.73--6.09) \\
&            & \textbf{16.27} & \textbf{68.00} & \textbf{11.27} & \textbf{4.45} \\
\midrule

\multirow{4}{*}{Alicante} 
& Indep       & \textcolor{red}{17.24 (13.73--20.85)} & \textcolor{red}{66.28 (62.20--69.83)} & 11.34 (8.64--14.41) & \textcolor{red}{5.14 (3.39--7.12)} \\
& Corr        & \textcolor{red}{16.71 (13.22--20.00)} & \textcolor{red}{66.87 (63.05--70.85)} & 11.35 (8.64--14.41) & \textcolor{red}{5.08 (3.22--7.12)} \\
& Corr\&IRE   & \cellcolor{gray!20}17.51 (14.58--21.02) & \textcolor{red}{63.87 (60.00--67.63)} & \cellcolor{gray!20}12.26 (9.49--14.92) & \cellcolor{gray!20}6.36 (4.75--8.14) \\
&            & \textbf{21.02} & \textbf{59.49} & \textbf{12.20} & \textbf{7.29} \\
\midrule

\multirow{4}{*}{Elche} 
& Indep       & \textcolor{red}{15.85 (12.59--19.31)} & \textcolor{red}{66.51 (62.41--70.69)} & 12.11 (9.14--15.34) & \textcolor{red}{5.53 (3.62--7.76)} \\
& Corr        & \textcolor{red}{15.89 (12.41--19.48)} & \textcolor{red}{66.75 (62.76--70.86)} & \cellcolor{gray!20}11.92 (8.97--15.00) & \textcolor{red}{5.44 (3.62--7.59)} \\
& Corr\&IRE   & \cellcolor{gray!20}17.06 (14.14--20.00) & \cellcolor{gray!20}63.01 (59.66--66.56) & 13.22 (10.69--15.87) & \cellcolor{gray!20}6.70 (4.83--8.62) \\
&            & \textbf{19.66} & \textbf{60.52} & \textbf{11.38} & \textbf{7.93} \\
\midrule

\multirow{4}{*}{\makecell{Castellón de\\la Plana}} 
& Indep       & 24.30 (18.08--31.15) & \textcolor{red}{64.55 (58.46--70.38)} & \textcolor{red}{7.79 (4.62--11.92)} & \textcolor{red}{3.36 (1.53--6.15)} \\
& Corr        & \cellcolor{gray!20}23.85 (18.45--30.00) & \textcolor{red}{64.95 (59.23--70.77)} & \textcolor{red}{7.90 (4.23--11.54)} & 3.30 (1.15--5.77) \\
& Corr\&IRE   & 23.93 (18.85--29.23) & \textcolor{red}{66.52 (61.15--71.92)} & \textcolor{red}{7.27 (4.23--10.38)} & \cellcolor{gray!20}2.28 (0.77--3.86) \\
&            & \textbf{22.31} & \textbf{73.08} & \textbf{3.46} & \textbf{1.15} \\
\midrule

\multirow{4}{*}{Orihuela} 
& Indep       & \cellcolor{gray!20}16.58 (11.46--22.40) & 66.53 (59.90--72.92) & \cellcolor{gray!20}11.64 (6.77--16.67) & 5.26 (2.08--8.35) \\
& Corr        & 17.77 (12.50--23.96) & \cellcolor{gray!20}66.94 (60.42--73.44) & 10.60 (6.25--15.10) & \cellcolor{gray!20}4.68 (2.07--8.33) \\
& Corr\&IRE   & 18.51 (13.54--23.44) & 64.41 (57.81--70.31) & 12.37 (8.33--17.71) & 4.71 (2.08--7.81) \\
&            & \textbf{16.15} & \textbf{68.75} & \textbf{11.46} & \textbf{3.12} \\
\midrule

\multirow{4}{*}{Torrevieja} 
& Indep       & 16.80 (10.87--23.37) & \textcolor{red}{66.48 (59.77--73.37)} & 11.48 (6.52--16.85) & 5.24 (2.17--9.24) \\
& Corr        & 16.93 (10.87--23.91) & \textcolor{red}{66.83 (60.33--73.37)} & 11.27 (6.52--16.85) & 4.97 (2.16--8.70) \\
& Corr\&IRE   & \cellcolor{gray!20}19.26 (14.13--25.00) & \cellcolor{gray!20}60.17 (53.80--66.30) & \cellcolor{gray!20}13.00 (8.70--17.93) & \cellcolor{gray!20}7.58 (4.35--10.87) \\
&            & \textbf{21.74} & \textbf{57.07} & \textbf{13.04} & \textbf{8.15} \\
\bottomrule
\end{tabular}
}
\vspace{0.25cm}
\caption{Model assessment for item 2 of the GHQ--12. Posterior means ($95\%$ prediction intervals) and observed values (in \textbf{bold}) of the percentage of respondents in each category across the selected municipalities in the RV. The model shaded in gray represents the best fit, as its prediction interval includes the observed value and its posterior mean is the closest among all models. Models whose prediction interval does not include the observed value are highlighted in red.}
\label{tab:assessment-2}
\end{center}
\end{table}

\begin{table}[!h]
\begin{center}
\resizebox{16.5cm}{!} {
\begin{tabular}{llllll}
\toprule
\multirow{2}{*}{\textbf{Municipality}} & \multirow{2}{*}{\textbf{Model}} & \multicolumn{4}{c}{\textbf{Response levels}} \\
\cmidrule(lr){3-6}
& & \textbf{More than usual} & \textbf{Same as usual} & \textbf{Less than usual} & \textbf{Much less than usual} \\
\midrule

\multirow{4}{*}{Valencia} 
& Indep       & \cellcolor{gray!20}7.07 (5.36--9.18) & 85.42 (82.91--87.73) & \cellcolor{gray!20}5.72 (4.36--7.36) & 1.79 (1.00--2.64) \\
& Corr        & 7.16 (5.45--9.00) & \cellcolor{gray!20}85.88 (83.73--87.91) & 5.32 (4.00--6.82) & \cellcolor{gray!20}1.63 (0.82--2.45) \\
& Corr\&IRE   & 7.21 (5.91--8.64) & 85.56 (83.64--87.18) & 5.50 (4.18--6.91) & 1.73 (1.09--2.45) \\
&            & \textbf{6.73} & \textbf{85.82} & \textbf{5.82} & \textbf{1.45} \\
\midrule

\multirow{4}{*}{Alicante} 
& Indep       & 9.56 (6.78--12.54) & 84.91 (82.03--87.97) & \cellcolor{gray!20}4.25 (2.54--6.10) & 1.29 (0.34--2.37) \\
& Corr        & 8.30 (5.76--11.19) & 85.64 (82.54--88.64) & 4.66 (2.88--6.61) & 1.40 (0.51--2.54) \\
& Corr\&IRE   & \cellcolor{gray!20}9.90 (7.63--12.37) & \cellcolor{gray!20}83.26 (80.51--85.93) & 4.94 (3.39--6.78) & \cellcolor{gray!20}1.90 (1.02--2.88) \\
&            & \textbf{11.02} & \textbf{82.54} & \textbf{4.41} & \textbf{2.03} \\
\midrule

\multirow{4}{*}{Elche} 
& Indep       & 11.82 (8.62--15.17) & 83.80 (80.69--87.24) & 3.38 (1.90--5.00) & \cellcolor{gray!20}1.00 (0.34--2.07) \\
& Corr        & 12.18 (9.31--15.17) & 83.81 (80.86--86.73) & 3.13 (1.72--4.66) & 0.89 (0.17--1.72) \\
& Corr\&IRE   & \cellcolor{gray!20}13.59 (11.55--15.86) & \cellcolor{gray!20}80.79 (78.28--83.28) & \cellcolor{gray!20}4.32 (2.93--6.03) & 1.31 (0.52--2.07) \\
&            & \textbf{13.10} & \textbf{81.38} & \textbf{3.97} & \textbf{1.03} \\
\midrule

\multirow{4}{*}{\makecell{Castellón de\\la Plana}} 
& Indep       & 6.94 (3.46--11.15) & 85.32 (80.38--89.62) & \cellcolor{gray!20}5.94 (3.08--9.62) & 1.81 (0.38--3.47) \\
& Corr        & 7.17 (3.85--10.38) & \cellcolor{gray!20}85.86 (81.54--90.00) & 5.35 (2.69--8.85) & 1.63 (0.38--3.46) \\
& Corr\&IRE   & \cellcolor{gray!20}6.06 (3.46--9.23) & 87.73 (84.23--91.15) & 5.14 (2.69--7.69) & \cellcolor{gray!20}1.08 (0.00--2.31) \\
&            & \textbf{6.15} & \textbf{85.77} & \textbf{6.54} & \textbf{1.15} \\
\midrule

\multirow{4}{*}{Orihuela} 
& Indep       & \cellcolor{gray!20}9.15 (4.69--14.08) & 84.92 (79.17--90.10) & \cellcolor{gray!20}4.57 (1.56--7.81) & 1.35 (0.00--3.12) \\
& Corr        & 10.79 (6.25--16.15) & \cellcolor{gray!20}84.63 (79.17--89.58) & 3.54 (1.04--6.25) & \cellcolor{gray!20}1.03 (0.00--2.60) \\
& Corr\&IRE   & 11.29 (7.81--15.10) & 82.79 (78.12--86.98) & 4.61 (2.08--7.29) & 1.31 (0.00--2.60) \\
&            & \textbf{9.90} & \textbf{84.38} & \textbf{4.17} & \textbf{1.04} \\
\midrule

\multirow{4}{*}{Torrevieja} 
& Indep       & 9.48 (4.89--14.67) & 84.69 (78.80--89.67) & 4.44 (1.63--7.61) & 1.40 (0.00--3.80) \\
& Corr        & \cellcolor{gray!20}10.38 (5.98--15.76) & \cellcolor{gray!20}84.73 (79.35--90.22) & \cellcolor{gray!20}3.75 (1.09--7.07) & 1.14 (0.00--2.72) \\
& Corr\&IRE   & 10.97 (7.61--14.67) & 83.64 (79.35--87.50) & 4.31 (1.63--7.07) & \cellcolor{gray!20}1.09 (0.00--2.72) \\
&            & \textbf{10.33} & \textbf{84.78} & \textbf{3.80} & \textbf{1.09} \\
\bottomrule
\end{tabular}
}
\vspace{0.25cm}
\caption{Model assessment for item 3 of the GHQ--12. Posterior means ($95\%$ prediction intervals) and observed values (in \textbf{bold}) of the percentage of respondents in each category across the selected municipalities in the RV. The model shaded in gray represents the best fit, as its prediction interval includes the observed value and its posterior mean is the closest among all models. Models whose prediction interval does not include the observed value are highlighted in red.}
\label{tab:assessment-3}
\end{center}
\end{table}

\begin{table}[!h]
\begin{center}
\resizebox{16.5cm}{!} {
\begin{tabular}{llllll}
\toprule
\multirow{2}{*}{\textbf{Municipality}} & \multirow{2}{*}{\textbf{Model}} & \multicolumn{4}{c}{\textbf{Response levels}} \\
\cmidrule(lr){3-6}
& & \textbf{More than usual} & \textbf{Same as usual} & \textbf{Less than usual} & \textbf{Much less than usual} \\
\midrule

\multirow{4}{*}{Valencia} 
& Indep     & 5.65 (4.09--7.36) & 87.66 (85.73--89.55) & \cellcolor{gray!20}5.01 (3.55--6.55) & 1.69 (0.91--2.55) \\
& Corr      & 5.46 (4.00--7.00) & 88.12 (86.09--90.00) & 4.80 (3.55--6.27) & 1.61 (0.91--2.45) \\
& Corr\&IRE & \cellcolor{gray!20}5.65 (4.64--6.82) & \cellcolor{gray!20}87.92 (86.27--89.55) & 4.86 (3.73--6.09) & \cellcolor{gray!20}1.56 (1.00--2.18) \\
&          & \textbf{5.73} & \textbf{87.91} & \textbf{5.18} & \textbf{1.18} \\
\midrule

\multirow{4}{*}{Alicante} 
& Indep     & 5.72 (3.56--8.14) & 87.68 (85.08--90.17) & 4.93 (3.05--7.12) & 1.67 (0.68--2.88) \\
& Corr      & 5.35 (3.39--7.46) & 88.11 (85.42--90.85) & \cellcolor{gray!20}4.91 (3.22--6.78) & 1.63 (0.68--2.71) \\
& Corr\&IRE & \cellcolor{gray!20}6.27 (4.58--7.97) & \cellcolor{gray!20}86.37 (84.07--88.48) & 5.10 (3.56--6.78) & \cellcolor{gray!20}2.26 (1.36--3.05) \\
&          & \textbf{6.44} & \textbf{86.10} & \textbf{4.75} & \textbf{2.71} \\
\midrule

\multirow{4}{*}{Elche} 
& Indep     & 9.19 (6.38--12.24) & 86.72 (83.62--89.66) & \cellcolor{gray!20}3.10 (1.72--4.66) & 0.99 (0.17--1.90) \\
& Corr      & 9.10 (6.38--11.90) & 87.07 (83.97--89.83) & 2.91 (1.55--4.48) & 0.92 (0.17--1.90) \\
& Corr\&IRE & \cellcolor{gray!20}10.95 (9.31--12.93) & \cellcolor{gray!20}83.88 (81.55--86.21) & 3.79 (2.41--5.35) & \cellcolor{gray!20}1.38 (0.69--2.07) \\
&          & \textbf{10.52} & \textbf{84.31} & \textbf{3.10} & \textbf{1.55} \\
\midrule

\multirow{4}{*}{\makecell{Castellón de\\la Plana}} 
& Indep     & \cellcolor{gray!20}6.35 (3.08--10.00) & \cellcolor{gray!20}87.69 (83.46--91.54) & \cellcolor{gray!20}4.49 (2.31--7.31) & 1.47 (0.00--3.08) \\
& Corr      & 5.98 (3.08--9.62) & 88.12 (83.85--91.92) & 4.46 (1.92--7.31) & 1.43 (0.38--3.08) \\
& Corr\&IRE & 5.10 (3.08--7.69) & 90.23 (87.30--93.08) & 3.85 (1.92--6.15) & \cellcolor{gray!20}0.82 (0.00--1.92) \\
&          & \textbf{6.92} & \textbf{87.31} & \textbf{5.38} & \textbf{0.38} \\
\midrule

\multirow{4}{*}{Orihuela} 
& Indep     & 7.36 (3.65--11.98) & 87.52 (82.29--92.19) & \cellcolor{gray!20}3.85 (1.04--6.77) & 1.28 (0.00--3.12) \\
& Corr      & 8.89 (4.69--13.54) & 87.19 (82.29--91.67) & 2.98 (1.04--5.73) & 0.94 (0.00--2.60) \\
& Corr\&IRE & \cellcolor{gray!20}10.00 (7.29--13.02) & \cellcolor{gray!20}84.54 (80.73--88.02) & 4.01 (1.56--6.77) & \cellcolor{gray!20}1.44 (0.52--2.60) \\
&          & \textbf{9.38} & \textbf{84.90} & \textbf{3.65} & \textbf{1.56} \\
\midrule

\multirow{4}{*}{Torrevieja} 
& Indep     & 8.56 (3.80--13.59) & 86.75 (81.52--91.85) & 3.50 (1.09--7.07) & 1.19 (0.00--2.72) \\
& Corr      & 8.44 (4.35--14.13) & \cellcolor{gray!20}87.27 (82.07--91.85) & \cellcolor{gray!20}3.23 (0.54--5.98) & 1.06 (0.00--2.72) \\
& Corr\&IRE & \cellcolor{gray!20}9.45 (7.07--12.50) & 85.79 (81.52--89.13) & 3.80 (1.63--6.52) & \cellcolor{gray!20}0.96 (0.00--2.17) \\
&          & \textbf{9.24} & \textbf{87.50} & \textbf{2.72} & \textbf{0.54} \\
\bottomrule
\end{tabular}
}
\vspace{0.25cm}
\caption{Model assessment for item 4 of the GHQ--12. Posterior means ($95\%$ prediction intervals) and observed values (in \textbf{bold}) of the percentage of respondents in each category across the selected municipalities in the RV. The model shaded in gray represents the best fit, as its prediction interval includes the observed value and its posterior mean is the closest among all models. Models whose prediction interval does not include the observed value are highlighted in red.}
\label{tab:assessment-4}
\end{center}
\end{table}

\begin{table}[!h]
\begin{center}
\resizebox{16.5cm}{!} {
\begin{tabular}{llllll}
\toprule
\multirow{2}{*}{\textbf{Municipality}} & \multirow{2}{*}{\textbf{Model}} & \multicolumn{4}{c}{\textbf{Response levels}} \\
\cmidrule(lr){3-6}
& & \textbf{Not at all} & \textbf{No more than usual} & \textbf{Rather more than usual} & \textbf{Much more than usual} \\

\midrule
\multirow{4}{*}{Valencia} 
& Indep       & 17.81 (15.18--20.55) & 56.02 (52.91--59.27) & 20.73 (18.00--23.73) & 5.43 (4.00--7.00) \\
& Corr       & \cellcolor{gray!20}17.18 (14.45--19.91) & 56.58 (53.45--59.82) & 20.91 (18.09--23.64) & 5.32 (4.00--6.82) \\
& Corr\&IRE & 17.41 (15.73--19.09) & \cellcolor{gray!20}57.31 (54.82--59.64) & \cellcolor{gray!20}20.50 (18.45--22.55) & \cellcolor{gray!20}4.78 (3.91--5.73) \\
&             & \textbf{16.91}       & \textbf{58.55}       & \textbf{20.18}       & \textbf{4.27} \\
\midrule
\multirow{4}{*}{Alicante} 
& Indep       & 15.44 (12.03--18.81) & 54.87 (50.34--59.15) & 23.25 (19.49--27.29) & \textcolor{red}{6.43 (4.41--8.81)} \\
& Corr       & 15.36 (12.37--18.64) & 55.48 (51.36--59.83) & 23.02 (19.15--26.61) & \textcolor{red}{6.14 (4.24--8.31)} \\
& Corr\&IRE & \cellcolor{gray!20}16.66 (14.41--18.98) & \cellcolor{gray!20}53.71 (50.68--56.95) & \cellcolor{gray!20}21.28 (18.47--24.07) & \cellcolor{gray!20}8.36 (6.95--9.83) \\
&             & \textbf{16.95}       & \textbf{53.39}       & \textbf{20.51}       & \textbf{9.15} \\
\midrule
\multirow{4}{*}{Elche} 
& Indep       & 14.83 (11.55--18.28) & \textcolor{red}{54.63 (50.51--58.79)} & 23.85 (20.00--28.11) & \textcolor{red}{6.68 (4.48--8.97)} \\
& Corr       & 14.57 (11.55--18.10) & \textcolor{red}{55.15 (50.86--59.48)} & 23.84 (19.82--27.76) & \textcolor{red}{6.44 (4.48--8.62)} \\
& Corr\&IRE & \cellcolor{gray!20}16.62 (14.48--18.79) & \cellcolor{gray!20}51.35 (48.28--54.31) & \cellcolor{gray!20}23.47 (20.69--26.03) & \cellcolor{gray!20}8.57 (7.07--10.17) \\
&             & \textbf{17.93}       & \textbf{50.00}       & \textbf{22.07}       & \textbf{9.48} \\
\midrule
\multirow{4}{*}{\makecell{Castellón de\\la Plana}}
& Indep       & 21.55 (15.77--28.08) & 56.64 (50.77--62.69) & 17.48 (12.31--23.09) & 4.33 (1.92--7.31) \\
& Corr       & 21.38 (15.77--27.31) & 57.23 (51.54--63.46) & 17.29 (12.31--22.69) & 4.10 (1.92--6.92) \\
& Corr\&IRE & \cellcolor{gray!20}20.02 (16.15--23.85) & \cellcolor{gray!20}60.72 (55.77--65.38) & \cellcolor{gray!20}16.68 (13.08--20.77) & \cellcolor{gray!20}2.58 (1.54--4.23) \\
&             & \textbf{20.00}       & \textbf{61.54}       & \textbf{16.15}       & \textbf{2.31} \\
\midrule
\multirow{4}{*}{Orihuela} 
& Indep       & 15.88 (10.42--21.89) & 55.09 (48.44--61.98) & 22.75 (16.15--29.69) & \textcolor{red}{6.28 (3.12--10.42)} \\
& Corr       & 17.37 (11.98--23.96) & 56.96 (50.63--65.55) & 20.50 (14.57--26.56) & 5.17 (2.08--8.85) \\
& Corr\&IRE & \cellcolor{gray!20}19.66 (16.15--23.44) & \cellcolor{gray!20}51.82 (46.88--56.77) & \cellcolor{gray!20}23.82 (19.27--28.12) & \cellcolor{gray!20}4.70 (2.60--7.29) \\
&             & \textbf{18.23}       & \textbf{52.08}       & \textbf{26.56}       & \textbf{2.60} \\
\midrule
\multirow{4}{*}{Torrevieja} 
& Indep       & 15.30 (9.78--21.21)  & 54.48 (47.28--61.96) & 23.56 (16.85--30.45) & 6.65 (3.25--10.87) \\
& Corr       & 15.49 (10.31--21.74) & 55.40 (47.83--63.04) & 23.03 (16.29--30.43) & 6.09 (2.72--10.33) \\
& Corr\&IRE & \cellcolor{gray!20}19.88 (16.30--23.37) & \cellcolor{gray!20}48.25 (43.48--53.80) & \cellcolor{gray!20}21.89 (17.39--26.63) & \cellcolor{gray!20}9.98 (7.61--12.50) \\
&             & \textbf{19.02}       & \textbf{49.46}       & \textbf{21.20}       & \textbf{10.33} \\
\bottomrule
\end{tabular}
}
\vspace{0.25cm}
\caption{Model assessment for item 5 of the GHQ--12. Posterior means ($95\%$ prediction intervals) and observed values (in \textbf{bold}) of the percentage of respondents in each category across the selected municipalities in the RV. The model shaded in gray represents the best fit, as its prediction interval includes the observed value and its posterior mean is the closest among all models. Models whose prediction interval does not include the observed value are highlighted in red.}
\label{tab:assessment-5}
\end{center}
\end{table}

\begin{table}[!h]
\begin{center}
\resizebox{16.5cm}{!} {
\begin{tabular}{llllll}
\toprule
\multirow{2}{*}{\textbf{Municipality}} & \multirow{2}{*}{\textbf{Model}} & \multicolumn{4}{c}{\textbf{Response levels}} \\
\cmidrule(lr){3-6}
& & \textbf{Not at all} & \textbf{No more than usual} & \textbf{Rather more than usual} & \textbf{Much more than usual} \\
\midrule

\multirow{4}{*}{Valencia} 
& Indep       & 12.18 (9.82--14.55) & 59.73 (56.36--62.82) & \cellcolor{gray!20}22.41 (19.55--25.45) & 5.68 (4.27--7.27) \\
& Corr        & 12.15 (9.91--14.27) & 60.47 (57.27--63.55) & 22.03 (19.36--24.82) & 5.35 (3.91--6.82) \\
& Corr\&IRE   & \cellcolor{gray!20}11.99 (10.45--13.55) & \cellcolor{gray!20}61.09 (58.55--63.45) & 21.80 (19.55--24.00) & \cellcolor{gray!20}5.12 (4.09--6.09) \\
&            & \textbf{10.82} & \textbf{62.36} & \textbf{22.27} & \textbf{4.55} \\
\midrule

\multirow{4}{*}{Alicante} 
& Indep       & \textcolor{red}{11.49 (8.64--14.75)} & \textcolor{red}{58.79 (54.24--63.22)} & 23.57 (19.66--27.97) & \textcolor{red}{6.14 (4.07--8.47)} \\
& Corr        & \textcolor{red}{10.78 (8.14--13.73)} & \textcolor{red}{58.99 (54.74--63.22)} & 24.13 (20.00--28.31) & \textcolor{red}{6.10 (4.07--8.31)} \\
& Corr\&IRE   & \textcolor{red}{12.30 (10.17--14.41)} & \textcolor{red}{56.58 (53.39--59.66)} & \cellcolor{gray!20}22.85 (19.83--25.93) & \textcolor{red}{8.26 (6.78--9.83)} \\
&            & \textbf{15.59} & \textbf{52.71} & \textbf{21.19} & \textbf{10.34} \\
\midrule

\multirow{4}{*}{Elche} 
& Indep       & 11.02 (8.27--14.14) & 58.67 (53.79--63.28) & 24.07 (19.83--28.45) & \textcolor{red}{6.25 (4.14--8.45)} \\
& Corr        & 10.32 (7.93--13.45) & 58.72 (54.31--63.10) & 24.65 (20.52--28.80) & \textcolor{red}{6.31 (4.31--8.62)} \\
& Corr\&IRE   & \cellcolor{gray!20}12.20 (10.00--14.48) & \cellcolor{gray!20}56.02 (52.93--59.66) & \cellcolor{gray!20}23.78 (21.03--26.90) & \cellcolor{gray!20}8.00 (6.55--9.66) \\
&            & \textbf{13.79} & \textbf{55.00} & \textbf{21.03} & \textbf{9.66} \\
\midrule

\multirow{4}{*}{\makecell{Castellón de\\la Plana}} 
& Indep       & 14.87 (10.00--20.38) & \textcolor{red}{61.38 (55.38--67.69)} & 19.14 (13.46--25.00) & \textcolor{red}{4.61 (2.31--7.69)} \\
& Corr        & 15.74 (10.77--21.15) & 62.55 (56.53--68.85) & \cellcolor{gray!20}17.74 (12.69--23.46) & 3.96 (1.54--6.92) \\
& Corr\&IRE   & \cellcolor{gray!20}13.61 (10.00--17.31) & \cellcolor{gray!20}66.41 (61.15--71.16) & 17.14 (13.08--21.54) & \cellcolor{gray!20}2.84 (1.54--4.62) \\
&            & \textbf{11.15} & \textbf{68.46} & \textbf{18.08} & \textbf{1.92} \\
\midrule

\multirow{4}{*}{Orihuela} 
& Indep       & 12.72 (7.29--18.23) & 60.05 (53.12--66.67) & 21.82 (15.09--28.14) & 5.41 (2.08--9.38) \\
& Corr        & 13.70 (8.33--19.27) & 61.31 (54.69--68.75) & 20.24 (14.06--26.56) & \cellcolor{gray!20}4.75 (2.08--8.33) \\
& Corr\&IRE   & \cellcolor{gray!20}16.63 (13.02--20.31) & \cellcolor{gray!20}55.59 (49.99--61.46) & \cellcolor{gray!20}22.94 (18.23--27.60) & 4.84 (2.60--7.29) \\
&            & \textbf{17.19} & \textbf{54.69} & \textbf{23.96} & \textbf{3.65} \\
\midrule

\multirow{4}{*}{Torrevieja} 
& Indep       & 12.51 (7.07--18.48) & \textcolor{red}{59.63 (52.72--66.85)} & 22.24 (15.22--29.90) & 5.62 (2.17--9.78) \\
& Corr        & \textcolor{red}{11.77 (7.07--17.39)} & \textcolor{red}{59.68 (52.17--66.85)} & 22.96 (16.83--29.89) & 5.60 (2.17--9.24) \\
& Corr\&IRE   & \cellcolor{gray!20}16.75 (13.04--20.65) & \cellcolor{gray!20}52.20 (46.73--57.61) & \cellcolor{gray!20}21.38 (16.30--26.63) & \cellcolor{gray!20}9.68 (7.07--12.50) \\
&            & \textbf{17.93} & \textbf{51.63} & \textbf{21.20} & \textbf{9.24} \\
\bottomrule
\end{tabular}
}
\vspace{0.25cm}
\caption{Model assessment for item 6 of the GHQ--12. Posterior means ($95\%$ prediction intervals) and observed values (in \textbf{bold}) of the percentage of respondents in each category across the selected municipalities in the RV. The model shaded in gray represents the best fit, as its prediction interval includes the observed value and its posterior mean is the closest among all models. Models whose prediction interval does not include the observed value are highlighted in red.}
\label{tab:assessment-6}
\end{center}
\end{table}

\begin{table}[!h]
\begin{center}
\resizebox{16.5cm}{!} {
\begin{tabular}{llllll}
\toprule
\multirow{2}{*}{\textbf{Municipality}} & \multirow{2}{*}{\textbf{Model}} & \multicolumn{4}{c}{\textbf{Response levels}} \\
\cmidrule(lr){3-6}
& & \textbf{More than usual} & \textbf{Same as usual} & \textbf{Less than usual} & \textbf{Much less than usual} \\
\midrule

\multirow{4}{*}{Valencia} 
& Indep       & 6.22 (4.64--8.00) & 83.61 (81.27--85.91) & 8.23 (6.36--10.18) & 1.93 (1.09--2.82) \\
& Corr        & \cellcolor{gray!20}5.84 (4.36--7.46) & 84.08 (81.82--86.45) & 8.19 (6.36--10.09) & 1.88 (1.09--2.73) \\
& Corr\&IRE   & 5.96 (5.00--7.09) & \cellcolor{gray!20}84.29 (82.36--86.18) & \cellcolor{gray!20}7.97 (6.55--9.45) & \cellcolor{gray!20}1.78 (1.09--2.45) \\
&            & \textbf{5.45} & \textbf{85.55} & \textbf{7.82} & \textbf{1.18} \\
\midrule

\multirow{4}{*}{Alicante} 
& Indep       & 5.39 (3.39--7.63) & 83.20 (80.00--86.44) & 9.24 (6.61--12.03) & 2.17 (1.02--3.56) \\
& Corr        & 5.02 (3.22--6.95) & 83.56 (80.17--86.61) & 9.26 (6.78--12.20) & 2.17 (1.02--3.39) \\
& Corr\&IRE   & \cellcolor{gray!20}5.75 (4.24--7.29) & \cellcolor{gray!20}82.14 (79.66--84.58) & \cellcolor{gray!20}9.37 (7.29--11.53) & \cellcolor{gray!20}2.73 (1.86--3.90) \\
&            & \textbf{6.27} & \textbf{81.19} & \textbf{9.83} & \textbf{2.54} \\
\midrule

\multirow{4}{*}{Elche} 
& Indep       & 9.05 (6.55--11.90) & 83.89 (80.69--86.90) & \cellcolor{gray!20}5.76 (3.79--8.10) & 1.30 (0.52--2.41) \\
& Corr        & 8.90 (6.38--11.72) & \textcolor{red}{84.46 (81.38--87.41)} & 5.47 (3.62--7.59) & 1.17 (0.34--2.24) \\
& Corr\&IRE   & \cellcolor{gray!20}10.73 (8.97--12.59) & \cellcolor{gray!20}80.82 (78.45--83.28) & 6.84 (5.00--8.62) & \cellcolor{gray!20}1.61 (0.86--2.42) \\
&            & \textbf{10.86} & \textbf{80.86} & \textbf{6.21} & \textbf{1.55} \\
\midrule

\multirow{4}{*}{\makecell{Castellón de\\la Plana}} 
& Indep       & 6.23 (3.08--9.62) & 83.63 (78.85--88.08) & 8.20 (4.62--12.31) & 1.94 (0.38--3.86) \\
& Corr        & 6.03 (3.08--9.23) & 84.24 (79.23--88.46) & 7.93 (4.23--11.92) & 1.80 (0.38--3.85) \\
& Corr\&IRE   & \cellcolor{gray!20}4.69 (2.31--7.31) & \cellcolor{gray!20}87.03 (83.46--90.38) & \cellcolor{gray!20}7.12 (4.23--10.00) & \cellcolor{gray!20}1.17 (0.38--2.31) \\
&            & \textbf{4.23} & \textbf{87.69} & \textbf{7.31} & \textbf{0.77} \\
\midrule

\multirow{4}{*}{Orihuela} 
& Indep       & 7.91 (4.17--12.50) & 84.02 (78.65--89.06) & 6.61 (3.12--10.94) & 1.46 (0.00--3.65) \\
& Corr        & 9.10 (5.21--14.06) & 84.40 (78.65--89.58) & \cellcolor{gray!20}5.35 (2.59--9.38) & \cellcolor{gray!20}1.14 (0.00--2.60) \\
& Corr\&IRE   & \cellcolor{gray!20}10.76 (7.81--14.06) & \cellcolor{gray!20}81.08 (76.56--85.42) & 6.68 (3.65--9.90) & 1.49 (0.52--3.12) \\
&            & \textbf{11.46} & \textbf{81.25} & \textbf{5.73} & \textbf{1.04} \\
\midrule

\multirow{4}{*}{Torrevieja} 
& Indep       & 6.41 (2.72--10.87) & 83.44 (77.72--88.59) & \cellcolor{gray!20}8.17 (3.80--13.04) & \cellcolor{gray!20}1.98 (0.00--4.35) \\
& Corr        & 7.77 (3.80--12.50) & 84.31 (78.79--89.67) & 6.47 (2.72--10.34) & 1.46 (0.00--3.26) \\
& Corr\&IRE   & \cellcolor{gray!20}8.81 (6.51--11.41) & \cellcolor{gray!20}81.15 (76.63--85.33) & 8.32 (4.89--11.96) & 1.72 (0.54--3.27) \\
&            & \textbf{8.15} & \textbf{80.98} & \textbf{7.61} & \textbf{3.26} \\
\bottomrule
\end{tabular}
}
\vspace{0.25cm}
\caption{Model assessment for item 7 of the GHQ--12. Posterior means ($95\%$ prediction intervals) and observed values (in \textbf{bold}) of the percentage of respondents in each category across the selected municipalities in the RV. The model shaded in gray represents the best fit, as its prediction interval includes the observed value and its posterior mean is the closest among all models. Models whose prediction interval does not include the observed value are highlighted in red.}
\label{tab:assessment-7}
\end{center}
\end{table}

\begin{table}[!h]
\begin{center}
\resizebox{16.5cm}{!} {
\begin{tabular}{llllll}
\toprule
\multirow{2}{*}{\textbf{Municipality}} & \multirow{2}{*}{\textbf{Model}} & \multicolumn{4}{c}{\textbf{Response levels}} \\
\cmidrule(lr){3-6}
& & \textbf{More than usual} & \textbf{Same as usual} & \textbf{Less than usual} & \textbf{Much less than usual} \\
\midrule

\multirow{4}{*}{Valencia} 
& Indep       & 4.95 (3.45--6.55) & 86.26 (84.09--88.27) & \cellcolor{gray!20}6.70 (5.18--8.37) & 2.09 (1.18--3.09) \\
& Corr        & 4.73 (3.27--6.18) & \cellcolor{gray!20}86.82 (84.82--89.00) & 6.45 (5.00--8.09) & 2.00 (1.18--3.00) \\
& Corr\&IRE   & \cellcolor{gray!20}4.81 (3.91--5.91) & 86.84 (85.27--88.45) & 6.41 (5.09--7.73) & \cellcolor{gray!20}1.94 (1.36--2.64) \\
&            & \textbf{4.82} & \textbf{86.73} & \textbf{6.73} & \textbf{1.73} \\
\midrule

\multirow{4}{*}{Alicante} 
& Indep       & \cellcolor{gray!20}4.32 (2.54--6.27) & 85.81 (82.88--88.64) & \cellcolor{gray!20}7.47 (5.08--10.00) & 2.41 (1.19--3.73) \\
& Corr        & 3.93 (2.37--5.76) & 86.14 (83.22--88.98) & 7.55 (5.42--10.00) & 2.38 (1.19--3.73) \\
& Corr\&IRE   & 4.43 (3.05--5.93) & \cellcolor{gray!20}85.02 (82.71--87.29) & 7.54 (5.76--9.49) & \cellcolor{gray!20}3.01 (2.03--4.24) \\
&            & \textbf{4.24} & \textbf{85.42} & \textbf{6.95} & \textbf{3.22} \\
\midrule

\multirow{4}{*}{Elche} 
& Indep       & 6.28 (4.14--8.62) & \textcolor{red}{86.77 (83.79--89.66)} & 5.31 (3.28--7.41) & 1.64 (0.69--2.76) \\
& Corr        & 6.74 (4.48--9.14) & \textcolor{red}{87.22 (84.31--89.83)} & 4.65 (2.93--6.55) & 1.39 (0.52--2.59) \\
& Corr\&IRE   & \cellcolor{gray!20}8.55 (7.06--10.17) & \cellcolor{gray!20}83.51 (81.21--85.69) & \cellcolor{gray!20}5.98 (4.31--7.76) & \cellcolor{gray!20}1.95 (1.21--2.76) \\
&            & \textbf{8.45} & \textbf{82.76} & \textbf{6.21} & \textbf{2.07} \\
\midrule

\multirow{4}{*}{\makecell{Castellón de\\la Plana}} 
& Indep       & 5.56 (2.69--8.85) & \textcolor{red}{86.55 (82.31--90.38)} & 6.01 (3.08--9.23) & 1.88 (0.38--3.85) \\
& Corr        & 5.24 (2.31--8.46) & \textcolor{red}{87.16 (82.69--91.15)} & 5.82 (3.08--9.23) & 1.78 (0.38--3.46) \\
& Corr\&IRE   & \cellcolor{gray!20}3.86 (1.92--6.15) & \cellcolor{gray!20}90.12 (86.92--93.08) & \cellcolor{gray!20}5.08 (2.69--7.69) & \cellcolor{gray!20}0.94 (0.00--1.93) \\
&            & \textbf{3.46} & \textbf{91.54} & \textbf{4.62} & \textbf{0.38} \\
\midrule

\multirow{4}{*}{Orihuela} 
& Indep       & 6.53 (3.12--10.94) & 86.75 (82.29--91.15) & \cellcolor{gray!20}5.19 (2.08--8.85) & 1.52 (0.00--3.65) \\
& Corr        & 8.08 (4.17--13.02) & 86.95 (81.77--91.67) & 3.81 (1.04--6.77) & \cellcolor{gray!20}1.16 (0.00--3.12) \\
& Corr\&IRE   & \cellcolor{gray!20}10.14 (7.29--13.02) & \cellcolor{gray!20}83.30 (79.69--86.98) & 4.93 (2.08--7.81) & 1.64 (0.52--3.12) \\
&            & \textbf{10.42} & \textbf{82.81} & \textbf{5.21} & \textbf{1.04} \\
\midrule

\multirow{4}{*}{Torrevieja} 
& Indep       & 7.13 (3.26--11.41) & 86.50 (80.98--91.85) & 4.85 (1.63--8.70) & 1.52 (0.00--3.80) \\
& Corr        & 7.15 (3.79--11.96) & \cellcolor{gray!20}87.08 (82.60--91.85) & \cellcolor{gray!20}4.40 (1.63--7.61) & 1.37 (0.00--3.26) \\
& Corr\&IRE   & \cellcolor{gray!20}8.50 (5.98--11.41) & 84.42 (80.43--88.04) & 5.72 (2.72--8.70) & \cellcolor{gray!20}1.36 (0.00--2.72) \\
&            & \textbf{8.15} & \textbf{86.96} & \textbf{4.35} & \textbf{0.54} \\
\bottomrule
\end{tabular}
}
\vspace{0.25cm}
\caption{Model assessment for item 8 of the GHQ--12. Posterior means ($95\%$ prediction intervals) and observed values (in \textbf{bold}) of the percentage of respondents in each category across the selected municipalities in the RV. The model shaded in gray represents the best fit, as its prediction interval includes the observed value and its posterior mean is the closest among all models. Models whose prediction interval does not include the observed value are highlighted in red.}
\label{tab:assessment-8}
\end{center}
\end{table}

\begin{table}[!h]
\begin{center}
\resizebox{16.5cm}{!} {
\begin{tabular}{llllll}
\toprule
\multirow{2}{*}{\textbf{Municipality}} & \multirow{2}{*}{\textbf{Model}} & \multicolumn{4}{c}{\textbf{Response levels}} \\
\cmidrule(lr){3-6}
& & \textbf{Not at all} & \textbf{No more than usual} & \textbf{Rather more than usual} & \textbf{Much more than usual} \\
\midrule

\multirow{4}{*}{Valencia} 
& Indep       & \textcolor{red}{17.02 (14.36--19.82)} & \textcolor{red}{51.56 (48.55--54.73)} & 21.85 (19.09--24.73) & \textcolor{red}{9.57 (7.64--11.73)} \\
& Corr        & \textcolor{red}{16.56 (14.18--19.09)} & \textcolor{red}{52.30 (49.09--55.73)} & 21.94 (19.36--24.73) & \textcolor{red}{9.20 (7.36--11.09)} \\
& Corr\&IRE   & \textcolor{red}{15.39 (13.91--16.91)} & \cellcolor{gray!20}54.61 (52.36--56.73) & \cellcolor{gray!20}22.11 (20.27--24.00) & \cellcolor{gray!20}7.89 (6.91--9.00) \\
&            & \textbf{13.82} & \textbf{56.64} & \textbf{22.45} & \textbf{7.00} \\
\midrule

\multirow{4}{*}{Alicante} 
& Indep       & \textcolor{red}{11.44 (8.81--14.41)} & \textcolor{red}{46.31 (41.52--51.36)} & \textcolor{red}{27.79 (23.39--32.04)} & \textcolor{red}{14.46 (11.02--17.97)} \\
& Corr        & \textcolor{red}{10.66 (8.14--13.39)} & \textcolor{red}{46.28 (41.53--50.85)} & \textcolor{red}{28.53 (24.58--32.54)} & \textcolor{red}{14.53 (11.36--17.97)} \\
& Corr\&IRE   & \textcolor{red}{14.27 (12.37--16.10)} & \cellcolor{gray!20}41.85 (39.15--44.58) & \textcolor{red}{24.23 (21.53--26.78)} & \textcolor{red}{19.65 (17.46--21.86)} \\
&            & \textbf{16.78} & \textbf{40.51} & \textbf{20.51} & \textbf{22.03} \\
\midrule

\multirow{4}{*}{Elche} 
& Indep       & \textcolor{red}{9.26 (6.55--12.07)} & 43.09 (38.10--48.11) & \textcolor{red}{30.27 (26.03--34.66)} & \textcolor{red}{17.38 (13.79--21.38)} \\
& Corr        & \textcolor{red}{9.32 (6.90--11.90)} & 44.06 (39.66--48.79) & \textcolor{red}{30.30 (26.21--34.31)} & \textcolor{red}{16.33 (12.93--19.83)} \\
& Corr\&IRE   & \cellcolor{gray!20}11.70 (10.00--13.45) & \cellcolor{gray!20}40.13 (37.59--42.59) & \textcolor{red}{26.90 (24.14--29.66)} & \textcolor{red}{21.27 (19.14--23.28)} \\
&            & \textbf{12.76} & \textbf{40.17} & \textbf{22.41} & \textbf{24.14} \\
\midrule

\multirow{4}{*}{\makecell{Castellón de\\la Plana}} 
& Indep       & 19.87 (14.23--25.78) & \textcolor{red}{52.78 (46.54--59.23)} & \cellcolor{gray!20}19.37 (14.23--25.00) & \textcolor{red}{7.98 (4.62--11.92)} \\
& Corr        & \textcolor{red}{20.19 (15.38--25.77)} & \textcolor{red}{53.71 (47.31--60.00)} & 18.75 (14.22--23.85) & \textcolor{red}{7.36 (4.23--10.77)} \\
& Corr\&IRE   & \cellcolor{gray!20}14.72 (11.54--18.08) & \cellcolor{gray!20}62.34 (57.69--66.54) & 19.18 (15.38--22.70) & \cellcolor{gray!20}3.76 (2.31--5.38) \\
&            & \textbf{14.62} & \textbf{62.31} & \textbf{19.62} & \textbf{3.46} \\
\midrule

\multirow{4}{*}{Orihuela} 
& Indep       & 20.78 (14.06--27.60) & 52.84 (45.83--59.38) & 18.78 (13.02--25.01) & 7.60 (3.65--11.98) \\
& Corr        & 21.78 (15.10--28.65) & 53.65 (46.88--60.42) & \textcolor{red}{17.82 (12.49--23.45)} & 6.75 (3.65--10.94) \\
& Corr\&IRE   & \cellcolor{gray!20}25.78 (22.40--29.17) & \cellcolor{gray!20}46.38 (41.67--51.04) & \cellcolor{gray!20}22.14 (18.23--26.04) & \cellcolor{gray!20}5.70 (3.65--7.81) \\
&            & \textbf{23.96} & \textbf{47.92} & \textbf{23.96} & \textbf{3.65} \\
\midrule

\multirow{4}{*}{Torrevieja} 
& Indep       & 17.55 (11.41--25.00) & 51.28 (44.02--58.15) & \cellcolor{gray!20}21.64 (14.67--28.80) & 9.53 (4.89--14.67) \\
& Corr        & 15.98 (10.33--22.28) & 51.53 (43.48--58.70) & 22.72 (16.30--29.89) & 9.77 (5.43--14.67) \\
& Corr\&IRE   & \cellcolor{gray!20}21.27 (17.93--24.46) & \cellcolor{gray!20}44.18 (39.67--48.91) & 21.34 (17.39--25.01) & \cellcolor{gray!20}13.21 (10.87--15.76) \\
&            & \textbf{20.65} & \textbf{45.65} & \textbf{21.74} & \textbf{11.96} \\
\bottomrule
\end{tabular}
}
\vspace{0.25cm}
\caption{Model assessment for item 9 of the GHQ--12. Posterior means ($95\%$ prediction intervals) and observed values (in \textbf{bold}) of the percentage of respondents in each category across the selected municipalities in the RV. The model shaded in gray represents the best fit, as its prediction interval includes the observed value and its posterior mean is the closest among all models. Models whose prediction interval does not include the observed value are highlighted in red.}
\label{tab:assessment-9}
\end{center}
\end{table}

\begin{table}[!h]
\begin{center}
\resizebox{16.5cm}{!} {
\begin{tabular}{llllll}
\toprule
\multirow{2}{*}{\textbf{Municipality}} & \multirow{2}{*}{\textbf{Model}} & \multicolumn{4}{c}{\textbf{Response levels}} \\
\cmidrule(lr){3-6}
& & \textbf{Not at all} & \textbf{No more than usual} & \textbf{Rather more than usual} & \textbf{Much more than usual} \\
\midrule

\multirow{4}{*}{Valencia} 
& Indep       & \textcolor{red}{15.45 (12.82--18.09)} & \textcolor{red}{53.71 (50.45--56.73)} & 21.53 (18.82--24.37) & \textcolor{red}{9.32 (7.36--11.27)} \\
& Corr        & \textcolor{red}{14.68 (12.54--17.09)} & \textcolor{red}{54.05 (50.91--57.27)} & \cellcolor{gray!20}21.95 (19.27--24.73) & \textcolor{red}{9.32 (7.45--11.27)} \\
& Corr\&IRE   & \cellcolor{gray!20}13.32 (12.09--14.55) & \textcolor{red}{56.76 (54.82--58.64)} & 22.12 (20.45--23.73) & \cellcolor{gray!20}7.80 (6.91--8.73) \\
&            & \textbf{12.36} & \textbf{58.82} & \textbf{21.82} & \textbf{7.00} \\
\midrule

\multirow{4}{*}{Alicante} 
& Indep       & \textcolor{red}{9.31 (6.95--11.87)} & 46.02 (41.19--50.68) & \textcolor{red}{28.83 (24.75--33.05)} & \textcolor{red}{15.85 (12.54--19.66)} \\
& Corr        & \textcolor{red}{8.60 (6.10--11.02)} & 45.78 (40.85--50.34) & \textcolor{red}{29.76 (25.59--34.07)} & \textcolor{red}{15.86 (12.54--19.49)} \\
& Corr\&IRE   & \cellcolor{gray!20}12.58 (11.19--14.24) & \cellcolor{gray!20}42.57 (40.17--44.92) & \textcolor{red}{23.11 (20.84--25.76)} & \textcolor{red}{21.74 (20.00--23.39)} \\
&            & \textbf{13.90} & \textbf{41.86} & \textbf{20.51} & \textbf{23.56} \\
\midrule

\multirow{4}{*}{Elche} 
& Indep       & \textcolor{red}{7.84 (5.51--10.52)} & 43.49 (38.45--48.28) & \textcolor{red}{30.65 (26.9--34.66)} & \textcolor{red}{18.01 (14.31--21.90)} \\
& Corr        & \textcolor{red}{7.65 (5.34--10.17)} & 43.81 (39.31--48.45) & \textcolor{red}{30.99 (26.9--35.00)} & \textcolor{red}{17.54 (13.97--21.21)} \\
& Corr\&IRE   & \cellcolor{gray!20}10.35 (8.97--11.72) & \cellcolor{gray!20}41.21 (38.96--43.45) & \cellcolor{gray!20}25.14 (22.59--27.93) & \cellcolor{gray!20}23.30 (21.38--25.17) \\
&            & \textbf{11.03} & \textbf{41.21} & \textbf{22.76} & \textbf{24.48} \\
\midrule

\multirow{4}{*}{\makecell{Castellón de\\la Plana}} 
& Indep       & \textcolor{red}{18.64 (13.08--24.24)} & \textcolor{red}{55.14 (48.85--61.15)} & 18.58 (13.46--24.23) & \textcolor{red}{7.63 (4.23--11.15)} \\
& Corr        & \textcolor{red}{18.18 (13.08--23.47)} & \textcolor{red}{55.98 (50.00--61.92)} & 18.56 (13.85--23.46) & \textcolor{red}{7.28 (4.23--10.77)} \\
& Corr\&IRE   & \cellcolor{gray!20}12.36 (9.62--15.00) & \textcolor{red}{66.08 (62.31--70.38)} & \cellcolor{gray!20}18.01 (14.62--21.92) & \cellcolor{gray!20}3.55 (2.31--5.00) \\
&            & \textbf{9.62} & \textbf{71.92} & \textbf{15.00} & \textbf{3.46} \\
\midrule

\multirow{4}{*}{Orihuela} 
& Indep       & 19.04 (12.50--25.53) & 55.14 (47.92--62.50) & 18.38 (12.5--25.00) & 7.44 (3.65--11.98) \\
& Corr        & 19.73 (13.02--26.04) & \textcolor{red}{56.30 (48.96--63.54)} & \textcolor{red}{17.41 (11.98--23.44)} & 6.57 (3.12--10.42) \\
& Corr\&IRE   & \cellcolor{gray!20}23.65 (20.83--27.08) & \cellcolor{gray!20}48.79 (44.27--53.12) & \cellcolor{gray!20}22.37 (18.23--26.04) & \cellcolor{gray!20}5.19 (3.65--7.29) \\
&            & \textbf{23.44} & \textbf{48.44} & \textbf{23.96} & \textbf{3.65} \\
\midrule

\multirow{4}{*}{Torrevieja} 
& Indep       & 14.85 (9.24--21.20) & 52.99 (45.65--60.33) & 22.10 (15.22--29.35) & \cellcolor{gray!20}10.06 (5.43--15.22) \\
& Corr        & 14.14 (8.70--20.11) & 53.26 (46.20--60.87) & \cellcolor{gray!20}22.73 (16.30--28.8) & 9.87 (5.98--14.67) \\
& Corr\&IRE   & \cellcolor{gray!20}18.97 (16.30--21.74) & \cellcolor{gray!20}45.54 (41.83--50.00) & 22.66 (19.01--26.63) & 12.83 (10.87--15.22) \\
&            & \textbf{18.48} & \textbf{46.20} & \textbf{23.91} & \textbf{11.41} \\
\bottomrule
\end{tabular}
}
\vspace{0.25cm}
\caption{Model assessment for item 10 of the GHQ--12. Posterior means ($95\%$ prediction intervals) and observed values (in \textbf{bold}) of the percentage of respondents in each category across the selected municipalities in the RV. The model shaded in gray represents the best fit, as its prediction interval includes the observed value and its posterior mean is the closest among all models. Models whose prediction interval does not include the observed value are highlighted in red.}
\label{tab:assessment-10}
\end{center}
\end{table}

\begin{table}[!h]
\begin{center}
\resizebox{16.5cm}{!} {
\begin{tabular}{llllll}
\toprule
\multirow{2}{*}{\textbf{Municipality}} & \multirow{2}{*}{\textbf{Model}} & \multicolumn{4}{c}{\textbf{Response levels}} \\
\cmidrule(lr){3-6}
& & \textbf{Not at all} & \textbf{No more than usual} & \textbf{Rather more than usual} & \textbf{Much more than usual} \\
\midrule

\multirow{4}{*}{Valencia} 
& Indep       & 11.54 (9.36--13.73) & 62.73 (59.73--65.73) & 15.04 (12.55--17.64) & \textcolor{red}{10.69 (8.63--13.00)} \\
& Corr        & \textcolor{red}{11.82 (9.73--14.27)} & 63.48 (60.45--66.46) & 14.59 (12.45--16.82) & 10.11 (8.18--12.27) \\
& Corr\&IRE   & \cellcolor{gray!20}10.66 (9.36--12.09) & \cellcolor{gray!20}64.87 (62.73--67.00) & \cellcolor{gray!20}15.60 (13.82--17.55) & \cellcolor{gray!20}8.87 (7.73--10.09) \\
&            & \textbf{9.64} & \textbf{65.45} & \textbf{16.64} & \textbf{8.27} \\
\midrule

\multirow{4}{*}{Alicante} 
& Indep       & \textcolor{red}{6.04 (4.07--8.14)} & \textcolor{red}{52.14 (46.78--57.46)} & \textcolor{red}{21.89 (18.31--25.76)} & \textcolor{red}{19.94 (15.93--24.24)} \\
& Corr        & \textcolor{red}{5.91 (3.90--7.97)} & \textcolor{red}{52.72 (47.79--56.95)} & \textcolor{red}{21.80 (18.47--25.42)} & \textcolor{red}{19.56 (15.76--23.56)} \\
& Corr\&IRE   & \textcolor{red}{9.06 (7.46--10.85)} & \textcolor{red}{49.19 (46.44--51.86)} & \cellcolor{gray!20}17.53 (14.92--20.34) & \cellcolor{gray!20}24.23 (22.37--26.27) \\
&            & \textbf{11.19} & \textbf{44.92} & \textbf{17.80} & \textbf{25.93} \\
\midrule

\multirow{4}{*}{Elche} 
& Indep       & \textcolor{red}{5.84 (3.79--8.10)} & 52.01 (47.07--57.24) & \textcolor{red}{22.05 (18.28--25.69)} & \textcolor{red}{20.10 (16.03--24.14)} \\
& Corr        & \textcolor{red}{5.42 (3.45--7.59)} & 51.62 (46.55--56.38) & \textcolor{red}{22.59 (19.14--26.21)} & \textcolor{red}{20.37 (16.38--24.31)} \\
& Corr\&IRE   & \textcolor{red}{8.32 (6.90--9.83)} & \cellcolor{gray!20}49.12 (46.38--51.90) & \cellcolor{gray!20}17.72 (15.17--20.52) & \cellcolor{gray!20}24.83 (22.76--26.90) \\
&            & \textbf{10.00} & \textbf{47.24} & \textbf{16.38} & \textbf{25.86} \\
\midrule

\multirow{4}{*}{\makecell{Castellón de\\la Plana}} 
& Indep       & \textcolor{red}{16.51 (11.15--22.31)} & \textcolor{red}{64.87 (58.85--70.38)} & \textcolor{red}{11.27 (7.30--15.77)} & 7.35 (4.23--11.16) \\
& Corr        & \textcolor{red}{16.46 (11.54--21.92)} & \textcolor{red}{65.28 (59.23--71.15)} & \textcolor{red}{11.06 (7.31--15.00)} & 7.20 (4.23--10.77) \\
& Corr\&IRE   & \textcolor{red}{11.11 (8.08--14.62)} & \textcolor{red}{75.36 (71.15--80.00)} & \textcolor{red}{9.92 (6.54--13.46)} & \cellcolor{gray!20}3.61 (2.31--5.00) \\
&            & \textbf{7.69} & \textbf{82.31} & \textbf{5.77} & \textbf{4.23} \\
\midrule

\multirow{4}{*}{Orihuela} 
& Indep       & 18.17 (11.98--25.00) & 64.92 (58.85--71.35) & 10.39 (6.25--15.10) & 6.51 (3.12--10.42) \\
& Corr        & 19.27 (13.02--25.53) & 65.12 (58.85--72.40) & 9.74 (5.73--14.06) & 5.88 (2.60--9.90) \\
& Corr\&IRE   & \cellcolor{gray!20}21.73 (17.71--25.53) & \cellcolor{gray!20}60.97 (55.73--65.62) & \cellcolor{gray!20}12.56 (8.33--16.67) & \cellcolor{gray!20}4.74 (3.12--6.77) \\
&            & \textbf{20.83} & \textbf{61.46} & \textbf{13.54} & \textbf{3.65} \\
\midrule

\multirow{4}{*}{Torrevieja} 
& Indep       & 11.90 (7.07--17.40) & \textcolor{red}{62.88 (55.43--70.12)} & 14.72 (9.24--20.67) & \cellcolor{gray!20}10.49 (5.98--16.30) \\
& Corr        & 12.24 (7.60--17.93) & \textcolor{red}{63.60 (56.51--70.65)} & 14.27 (9.24--20.65) & 9.89 (5.43--14.67) \\
& Corr\&IRE   & \cellcolor{gray!20}16.72 (13.04--20.11) & \cellcolor{gray!20}54.72 (50.00--59.78) & \cellcolor{gray!20}16.03 (11.96--20.65) & 12.53 (9.78--15.76) \\
&            & \textbf{17.39} & \textbf{50.54} & \textbf{20.65} & \textbf{11.41} \\
\bottomrule
\end{tabular}
}
\vspace{0.25cm}
\caption{Model assessment for item 11 of the GHQ--12. Posterior means ($95\%$ prediction intervals) and observed values (in \textbf{bold}) of the percentage of respondents in each category across the selected municipalities in the RV. The model shaded in gray represents the best fit, as its prediction interval includes the observed value and its posterior mean is the closest among all models. Models whose prediction interval does not include the observed value are highlighted in red.}
\label{tab:assessment-11}
\end{center}
\end{table}

\begin{table}[!h]
\begin{center}
\resizebox{16.5cm}{!} {
\begin{tabular}{llllll}
\toprule
\multirow{2}{*}{\textbf{Municipality}} & \multirow{2}{*}{\textbf{Model}} & \multicolumn{4}{c}{\textbf{Response levels}} \\
\cmidrule(lr){3-6}
& & \textbf{More than usual} & \textbf{Same as usual} & \textbf{Less than usual} & \textbf{Much less than usual} \\
\midrule

\multirow{4}{*}{Valencia} 
& Indep       & 7.23 (5.55--9.00) & 85.42 (83.27--87.45) & 6.11 (4.73--7.82) & 1.24 (0.64--2.00) \\
& Corr        & 7.16 (5.36--9.09) & 85.80 (83.55--88.00) & 5.87 (4.45--7.36) & \cellcolor{gray!20}1.17 (0.55--1.91) \\
& Corr\&IRE   & \cellcolor{gray!20}7.10 (5.82--8.45) & \cellcolor{gray!20}85.96 (84.18--87.73) & \cellcolor{gray!20}5.76 (4.45--7.18) & 1.18 (0.64--1.82) \\
&            & \textbf{6.64} & \textbf{86.73} & \textbf{5.73} & \textbf{0.91} \\
\midrule

\multirow{4}{*}{Alicante} 
& Indep       & \cellcolor{gray!20}6.56 (4.24--9.15) & 85.36 (82.37--88.31) & 6.69 (4.41--9.15) & 1.38 (0.51--2.54) \\
& Corr        & 6.34 (4.24--8.64) & 85.80 (82.88--88.48) & 6.53 (4.24--8.81) & \cellcolor{gray!20}1.34 (0.51--2.37) \\
& Corr\&IRE   & 7.17 (5.25--9.33) & \cellcolor{gray!20}84.25 (81.69--86.78) & \cellcolor{gray!20}6.91 (5.08--8.98) & 1.66 (0.85--2.71) \\
&            & \textbf{6.61} & \textbf{84.58} & \textbf{7.29} & \textbf{1.36} \\
\midrule

\multirow{4}{*}{Elche} 
& Indep       & 9.94 (7.07--12.93) & \textcolor{red}{84.75 (81.72--87.76)} & 4.44 (2.75--6.55) & 0.87 (0.17--1.72) \\
& Corr        & 10.31 (7.58--13.62) & \textcolor{red}{84.89 (81.55--87.76)} & 4.02 (2.59--5.69) & 0.77 (0.17--1.55) \\
& Corr\&IRE   & \cellcolor{gray!20}11.67 (9.48--13.79) & \cellcolor{gray!20}82.06 (79.31--84.66) & \cellcolor{gray!20}5.20 (3.62--7.07) & \cellcolor{gray!20}1.08 (0.34--1.90) \\
&            & \textbf{12.59} & \textbf{80.00} & \textbf{5.52} & \textbf{1.38} \\
\midrule

\multirow{4}{*}{\makecell{Castellón de\\la Plana}} 
& Indep       & 6.76 (3.46--10.77) & \textcolor{red}{85.34 (80.77--89.62)} & 6.59 (3.45--10.38) & \cellcolor{gray!20}1.31 (0.00--2.69) \\
& Corr        & 6.84 (3.85--10.38) & 85.79 (81.15--90.00) & 6.19 (3.08--9.62) & 1.19 (0.00--2.69) \\
& Corr\&IRE   & \cellcolor{gray!20}5.28 (2.69--8.08) & \cellcolor{gray!20}88.07 (84.23--91.54) & \cellcolor{gray!20}5.85 (3.46--8.85) & 0.80 (0.00--1.92) \\
&            & \textbf{4.23} & \textbf{90.00} & \textbf{4.23} & \textbf{1.54} \\
\midrule

\multirow{4}{*}{Orihuela} 
& Indep       & 8.71 (4.17--13.54) & 85.17 (79.69--90.10) & 5.18 (2.08--8.85) & 0.95 (0.00--2.60) \\
& Corr        & 10.18 (5.73--15.10) & 84.89 (79.69--89.58) & 4.17 (1.56--7.29) & \cellcolor{gray!20}0.76 (0.00--2.08) \\
& Corr\&IRE   & \cellcolor{gray!20}11.10 (7.29--14.58) & \cellcolor{gray!20}82.75 (78.12--86.99) & \cellcolor{gray!20}5.25 (2.60--8.33) & 0.91 (0.00--2.60) \\
&            & \textbf{12.50} & \textbf{80.73} & \textbf{5.73} & \textbf{0.52} \\
\midrule

\multirow{4}{*}{Torrevieja} 
& Indep       & 8.31 (3.80--13.59) & \cellcolor{gray!20}84.98 (79.35--90.22) & 5.52 (2.17--9.24) & \cellcolor{gray!20}1.19 (0.00--2.72) \\
& Corr        & \cellcolor{gray!20}9.00 (4.89--14.13) & 85.24 (79.89--90.23) & \cellcolor{gray!20}4.75 (1.63--8.15) & 1.00 (0.00--2.72) \\
& Corr\&IRE   & 9.71 (6.52--13.04) & 83.35 (79.35--87.50) & 5.83 (2.72--9.24) & 1.11 (0.00--2.19) \\
&            & \textbf{9.24} & \textbf{84.24} & \textbf{4.89} & \textbf{1.63} \\
\bottomrule
\end{tabular}
}
\vspace{0.25cm}
\caption{Model assessment for item 12 of the GHQ--12. Posterior means ($95\%$ prediction intervals) and observed values (in \textbf{bold}) of the percentage of respondents in each category across the selected municipalities in the RV. The model shaded in gray represents the best fit, as its prediction interval includes the observed value and its posterior mean is the closest among all models. Models whose prediction interval does not include the observed value are highlighted in red.}
\label{tab:assessment-12}
\end{center}
\end{table}

\end{document}